\documentclass[aps,prb,reprint,superscriptaddress]{revtex4-1}
\usepackage[english]{babel}
\usepackage[T1]{fontenc}
\usepackage[utf8]{inputenc}

\usepackage{bm}
\usepackage{xcolor}
\usepackage{tabularx}
\usepackage{amssymb}
\usepackage{amsmath}
\usepackage{graphicx}
\usepackage{multirow}

\usepackage[pdftex]{hyperref}

\newcommand{\kk}{\mathbf{k}}
\newcommand{\q}{\mathbf{q}}
\newcommand{\w}{\omega}
\newcommand{\rr}{\mathbf{r}}
\newcommand{\dif}{\mathrm{d}}
\newcommand{\ep}{\epsilon}
\newcommand{\E}{\mathbf{E}}
\newcommand{\B}{\mathbf{B}}
\newcommand{\J}{\mathbf{J}}
\newcommand{\Curl}{\nabla\times}
\newcommand{\Div}{\nabla\cdot}

\newcommand{\etal}{\emph{et al. }}
\newcommand{\un}[1]{\mathrm{\ #1}}

\newcommand{\mob}{\mathrm{cm^2\, V^{-1}\, s^{-1}}}
\newcommand{\cmm}{\mathrm{cm^{-3}}}
\newcommand{\mc}[1]{\multicolumn{2}{l}{#1}}
\newcommand{\mr}[1]{\multirow{2}{*}{#1}}

\begin{document}

\title{Two-fluid hydrodynamic model for semiconductors}
\author{Johan~R.~Maack}
\affiliation{Department of Photonics Engineering, Technical University of Denmark, \O rsteds Plads 343, DK-2800~Kongens~Lyngby, Denmark}
\author{N.~Asger~Mortensen}
\affiliation{Center for Nano Optics \& Danish Institute for Advanced Study, University of Southern Denmark, Campusvej 55, DK-5230~Odense~M, Denmark}
\author{Martijn~Wubs}
\affiliation{Department of Photonics Engineering, Technical University of Denmark, \O rsteds Plads 343, DK-2800~Kongens~Lyngby, Denmark}

\date{\today}

\begin{abstract}
The hydrodynamic Drude model (HDM) has been successful in describing the optical properties of metallic nanostructures, but for semiconductors where several different kinds of charge carriers are present, an extended theory is required. We present a two-fluid hydrodynamic model for semiconductors containing electrons and holes (from thermal or external excitation) or light and heavy holes (in $p$-doped materials). The two-fluid model predicts the existence of two longitudinal modes, an acoustic and an optical, whereas only an optical mode is present in the HDM. By extending nonlocal Mie theory to two plasmas, we are able to simulate the optical properties of two-fluid nanospheres and predict that the acoustic mode gives rise to peaks in the extinction spectra that are absent in the HDM.\end{abstract}

%\pacs{78.20.Bh, 73.20.Mf, 78.30.Fs}

\maketitle

\section{Introduction}
The study of collective excitations of electrons, plasmonics, takes place on ever smaller scales as the fabrication and characterization techniques continue to improve. While this allows for the design of entirely new devices and materials with promising properties, it also requires improved theoretical tools to properly model the systems.

Metals, by far the most widely used plasmonic materials, are often described very accurately by the Drude model. But when the sizes approach the nanoscale, the model is no longer able to explain experimentally observable phenomena like, for example, the blueshift of the resonance frequency of the localized surface plasmon (LSP) in metallic nanospheres.\cite{tiggesbaumker93} An improved model that has been successful in describing the optical properties of metals on the nanoscale is the hydrodynamic Drude model (HDM).\cite{vielma07,abajo08,mcmahon09,raza13,deceglia13,christensen14,scalora14,raza15,toscano15,ciraci16,fitzgerald16} In this model, the polarization depends \emph{nonlocally} on the electrical field, and the aforementioned blueshift appears as a size-dependent nonlocal effect.\cite{raza13b,christensen14,raza15b} Furthermore, the HDM also predicts the existence of confined bulk plasmons in nanoparticles,\cite{raza13,christensen14} something that also has been found experimentally.\cite{lindau71}

While metals are the most commonly used plasmonic materials because of their large density of free electrons, semiconductors are also interesting due to the tunability of the electron density, either statically by doping or dynamically by applying a bias. Furthermore, intrinsic semiconductors may contain plasmas created either thermally or by external excitations (e.g. from a laser), and here the electron density can be controlled dynamically with the temperature or the excitation energy, respectively. Plasmonics has already been shown in several papers for doped semiconductors,\cite{anderson71,matz81,grychowski86,betti89,meng91,biagi92,ginn11,sachet15,luther11,garcia11,manthiram12,schimpf13,zhou15} biased semiconductors,\cite{liu17,liu18,dionne09,anglin11} laser excited semiconductors\cite{yang17} and thermally excited intrinsic semiconductors.\cite{bell98,rivas06,isaac08,hanham12}

Among these studies, Refs.~\onlinecite{luther11,garcia11,manthiram12,schimpf13,zhou15,liu17,liu18} investigated plasmons in nanostructures of semiconductors, but except for Refs.~\onlinecite{schimpf13} and \onlinecite{liu17} they all used the Drude model to describe their results. And just as for metals, one would expect that the Drude model only is accurate for semiconductor structures down to a certain size. Now, it is well known that semiconductor particles of only a few nanometers behave as quantum dots, but in the intermediate size regime between structures described by the Drude model and quantum dots, a different theoretical framework is needed (see e.g. Refs.~\onlinecite{hapala13,zhang14,carmina15}). 

Recently we made a case for applying the HDM to semiconductor structures in the mentioned intermediate size regime.\cite{maack17} In that study, we adapted the HDM to nanospheres made of doped semiconductors and intrinsic semiconductors with thermally excited charge carriers. In both cases we found that the nonlocal blueshift was even more pronounced than in metals and occurred in larger particles. In essence, this can be attributed to the increased Fermi wavelength and smaller effective mass in semiconductors, as compared to that in metals. Based on the HDM we also predicted the existence of standing bulk plasmons above the plasma frequency in semiconductors, and very recently these resonances were measured by De Ceglia \etal\cite{deceglia17a} in doped semiconductors. These interesting new developments are no doubt only the beginning of a series of investigations of hydrodynamic behavior in various semiconductor structures. 

In the present study, we propose an extension of the HDM for semiconductors. In Ref.~\onlinecite{maack17} we assumed that only electrons were present as charge carriers (and so did De Ceglia \etal\cite{deceglia17a}), and due to the generally smaller effective mass of the electrons compared to the holes, this is a reasonable approximation whenever electrons are present as majority charge carriers. In general, however, semiconductors may contain several different kinds of charge carriers such as electrons, heavy holes and light holes, and ideally all should be taken into account. Therefore, the aim of this paper is to develop a hydrodynamic model for materials containing more than one kind of charge carrier. We will restrict ourselves to include only two different types of charge carriers, e.g. electrons and holes or heavy and light holes, and call the model the \emph{hydrodynamic two-fluid model} (as opposed to the HDM which contains only one hydrodynamic fluid). Other models that include multiple charge carriers already exist in the form of transport equations\cite{jungel,sze} and quantum mechanical and semi-classical theories.\cite{pines56,entinwohlman84,schaefer87,scott93,bonitz00,zhang17} And while Ref.~\onlinecite{schaefer87} briefly considers the hydrodynamic model for a two-fluid system, we will here present a more detailed analysis of the optical properties. We will also consider finite systems which, to our knowledge, has not been done before. Our extension of a single fluid (appropriate for majority-carrier systems) to a two-fluid description shows interesting phenomena beyond the independent-fluids approximation that constitute an integral part of the local-response electrodynamics of doped semiconductors, i.e. the mere addition of electron and hole conductivities.\cite{sze}

In the next section, we will present the theoretical foundation for the two-fluid model, which will then be supported by a microscopic derivation in section~\ref{micro}. In section~\ref{bulk} we will discuss some of the general properties of the model, while in section~\ref{mie} we will focus on systems of spherical geometry and derive extended versions of the Mie-coefficients that take two hydrodynamic fluids into account. These coefficients will be used in section~\ref{results} where the optical properties of semiconductor nanoparticles will be calculated.

\section{The model}
\label{twofluid}

In the traditional HDM, the electrical field and the current density are determined by a wave equation and a hydrodynamic equation-of-motion.\cite{raza15} A natural extension to the HDM is therefore to include multiple hydrodynamic plasmas, each described with a hydrodynamic equation-of-motion. In the model presented here, we will consider two different kinds of charge carriers (or fluids), such as electrons and holes or light and heavy holes. The governing equations for the two-fluid model are therefore

\begin{subequations}
\label{eq:hydro2}
\begin{align}
\label{eq:hydroa}
\frac{\beta_a^2}{\w^2+i\gamma_a\w}\nabla\left(\nabla\cdot\J_a\right)+ \J_a=\frac{i\w\ep_0\w_{a}^2}{\w^2+i\gamma_a\w}\E, \\
\label{eq:hydrob}
\frac{\beta_b^2}{\w^2+i\gamma_b\w}\nabla\left(\nabla\cdot\J_b\right)+ \J_b=\frac{i\w\ep_0\w_{b}^2}{\w^2+i\gamma_b\w}\E, \\
\label{eq:wave2}
-\nabla\times\nabla\times\E+ \frac{\w^2}{c^2}\ep_{\infty}\E=-i\mu_0\w\left(\J_a+\J_b\right), 
\end{align}
\end{subequations}
where (\ref{eq:hydroa}) and (\ref{eq:hydrob}) are the linearized hydrodynamic equations of motion related to the charge carriers $a$ and $b$, respectively, and (\ref{eq:wave2}) is the wave equation originating from Maxwell's Equations. Here $\w_a$ and $\w_b$ are the plasma frequencies for the two fluids,  $\gamma_a$ and $\gamma_b$ are the damping constants, and $\beta_a$ and $\beta_b$ are the nonlocal parameters. Note, that if one of the current densities is set to zero (whereby the corresponding hydrodynamic equation can be removed), the equations reduce to the original equations of the HDM (see Eqs.~(15) in Ref.~\onlinecite{raza15}). Although not considered here, it is also clear that the model easily could be extended to more than two types of charge carriers. 

The real-space equations will be the starting point for most practical problems, but it can also be instructive to look in the reciprocal space as well. If the material is assumed to be infinite, the spacial Fourier transforms of Eqs.~(\ref{eq:hydro2}) are\cite{wubs15}
\begin{align}
\label{eq:hydroi_ft}
-\frac{\beta_i^2}{\w^2+i\gamma_i\w}\q\left(\q\cdot\J_i\right)+ \J_i=\frac{i\w\ep_0\w_i^2}{\w^2+i\gamma_i\w}\E ,\\
\label{eq:wave2_ft}
\q\times\q\times\E+ \frac{\w^2}{c^2}\ep_{\infty}\E=-i\mu_0\w\left(\J_a+\J_b\right),
\end{align}
where $\q$ is the wave vector and $i=a,b$. Let us now consider the transversal and the longitudinal parts of the field separately. Starting with the transversal, or divergence-free, part of the field, this has the property $\q\cdot\E^T=0$ (and similarly for $\J_a^T$ and $\J_b^T$). This also means that $\q\times\q\times\E^T=-q^2\E^T$, and Eqs.~(\ref{eq:wave2_ft}) and (\ref{eq:hydroi_ft}) can be combined to
\begin{equation}
\label{eq:ep_trans1}
q^2=\left(\ep_{\infty}-\frac{\w_a^2}{\w^2+i\gamma_a\w}-\frac{\w_b^2}{\w^2+i\gamma_b\w}\right)\frac{\w^2}{c^2} .
\end{equation}
From the relation $\ep_T\w^2/c^2=q^2$ we now see that the transversal dielectric function is given by
\begin{equation}
\label{eq:ep_trans2}
\ep_T(\w)=\ep_{\infty}-\frac{\w_a^2}{\w^2+i\gamma_a\w}-\frac{\w_b^2}{\w^2+i\gamma_b\w} .
\end{equation} 

The longitudinal, or rotation-free, part of the field has the property $\q\times\E^L=0$ (and similarly for $\J_a^L$ and $\J_b^L$). This means that $\q\left(\q\cdot\J_i^L\right)=q^2\J_i^L$, and Eqs.~(\ref{eq:wave2_ft}) and (\ref{eq:hydroi_ft}) give us
\begin{equation}
\label{eq:ep_long1}
0=\ep_{\infty}-\frac{\w_a^2}{\w^2+i\gamma_a\w-\beta_a^2q^2}-\frac{\w_b^2}{\w^2+i\gamma_b\w-\beta_b^2q^2} .
\end{equation}
From the relation $\ep_L=0$ we now see that the longitudinal dielectric function is given by
\begin{equation}
\label{eq:ep_long2}
\ep_L(q,\w)=\ep_{\infty}-\frac{\w_a^2}{\w^2+i\gamma_a\w-\beta_a^2q^2}-\frac{\w_b^2}{\w^2+i\gamma_b\w-\beta_b^2q^2} .
\end{equation}

We here see that $\ep_L$ is nonlocal (i.e. depends on the wavenumber $q$), while $\ep_T$ is local. This can be compared with the dielectric functions of the single-fluid HDM\cite{wubs15}
\begin{subequations}
\begin{align}
\label{eq:ep_trans3}
\ep_T(\w)&=\ep_{\infty}-\frac{\w_p^2}{\w^2+i\gamma\w} ,\\
\label{eq:ep_long3}
\ep_L(q,\w)&=\ep_{\infty}-\frac{\w_p^2}{\w^2+i\gamma\w-\beta^2q^2} ,
\end{align}
\end{subequations}
where $\ep_T$ and $\ep_L$ also are local and nonlocal, respectively.

For the two-fluid model, we notice that if the fluids have the same $\gamma$'s and $\beta$'s, then the plasma frequencies in the nominators of Eqs.~(\ref{eq:ep_trans2}) and (\ref{eq:ep_long2}) could be combined into a single effective parameter given by 
\begin{equation}
\label{eq:wp}
\w_{\rm eff}^2=\w_a^2+\w_b^2 ,
\end{equation}
whereby the expressions for $\ep_T$ and $\ep_L$ become equal to Eqs.~(\ref{eq:ep_trans3}) and (\ref{eq:ep_long3}), respectively. In other words, a two-fluid system can effectively be described by the single-fluid HDM whenever both $\gamma_a=\gamma_b$ and $\beta_a=\beta_b$.

\section{Microscopical foundation}
\label{micro}

In this section, we will show that the expression for $\ep_L$ in the two-fluid model in fact can be derived from quantum mechanics by using a slightly modified version of the Lindhard approximation. We will consider a system of fermions described by the Hamiltonian $H_0$ subject to a perturbation of the form
\begin{equation}
\label{eq:hamilton}
H_1=U_0e^{i\left(\q\cdot\rr-\w t\right)}+U_0^*e^{-i\left(\q\cdot\rr-\w t\right)} ,
\end{equation}
where $U_0$ is the amplitude of the perturbation. According to Fermi's Golden Rule, this results in the following expression for the longitudinal dielectric function\cite{giuseppe}
\begin{align}
\label{eq:golden}
&\ep_L(\q,\w)= \nonumber \\
&1+\frac{2 e^2}{\ep_0 q^2}\frac{1}{V}\sum_{\alpha\beta}\frac{|\langle\psi_\beta |e^{i\q\cdot\rr}|\psi_\alpha\rangle|^2}{E_\beta-E_\alpha -\hbar\w-i\eta} \lbrack f(E_\alpha)-f(E_\beta)\rbrack ,
\end{align}
where $V$ is the volume, $e$ is the elementary charge and $E$ is the energy (the electrical field is not used in this section so there is no risk of confusion). The excitation takes place between the states $|\psi_\alpha\rangle$ and $|\psi_\beta\rangle$, the function $f$ is the Fermi--Dirac distribution, and $\eta$ is a small real number originating from the Dirac identity.\cite{giuseppe}

We will now apply the Lindhard approximation in which the bands are assumed to be isotropic and perfectly parabolic and the wavefunctions are plane waves. This means that the matrix element in $\ep_L(\q,\w)$ equals 1 when the excitation is from $\kk$ to $\kk+\q$ and zero otherwise. But different from the typical Lindhard approximation in which only a single band is taken into account, we will here include two bands in the derivation. Excitations between these two bands are neglected, however, which is a reasonable approximation when considering energies smaller than the bandgap. The result is
\begin{equation}
\label{eq:ep_long4}
\ep_L(\q,\w)=1+\chi_a(\q,\w)+\chi_b(\q,\w) ,
\end{equation}
where the susceptibilities for bands $a$ and $b$ are given by
\begin{equation}
\label{eq:chi1}
\chi_i(\q,\w)=\frac{2 e^2}{\ep_0 q^2}\frac{1}{V}\sum_{\kk}\frac{f_i(\kk)-f_i(\kk+\q)}{E_i(\kk+\q)-E_i(\kk) -\hbar\w-i\eta} .
\end{equation}

In Appendix~\ref{dielectric} we show that in the $\q\to\mathrm{0}$ limit, Eq.~(\ref{eq:chi1}) can be rewritten as
\begin{equation}
\label{eq:chi2}
\chi_i(q,\w)=-\frac{\w_i^2}{\w^2+i\gamma\w}-\frac{\w_i^2}{(\w^2+i\gamma\w)^2}\beta_i^2q^2-\ldots ,
\end{equation}
where $\gamma$ is the damping constant and the plasma frequencies are given by
\begin{equation}
\label{eq:plasma}
\w_i^2=\frac{e^2 n_i}{\ep_0 m_i^*} . \\
\end{equation}
Here $n_i$ and $m_i^*$ are the charge carrier density and the effective mass, respectively, of band $i$. The nonlocal parameter $\beta_i$ depends on the nature of the charge carriers. In this paper we will consider them to be electrons and holes in an intrinsic semiconductor originating either from thermal excitation or laser excitation across the band gap, or heavy and light holes in a $p$-doped semiconductor. As shown in Appendix~\ref{dielectric}, the nonlocal parameter is in these cases given by
\begin{align}
\label{eq:beta1}
&\left.
\begin{array}{l}
\text{Thermally excited} \\
\quad\text{ch. carriers}
\end{array} \right. &\beta_i^2 &=\frac{3 k_B T}{ m_i^*} , \\
\label{eq:beta2}
&\left.
\begin{array}{l}
\text{Laser excited ch. carriers} \\
\text{Heavy and light holes}
\end{array} \right. & \beta_i^2 &=\frac{3 k_{Fi}^2\hbar^2}{5  m_i^{*2}}=\frac{3 }{5}v_{Fi}^2 , 
\end{align}
where $T$ is the temperature, $k_B$ is Boltzmann's constant, and $k_{Fi}$ and $v_{Fi}$ are the Fermi wavenumber and the Fermi velocity, respectively, of band $i$. For thermally excited charge carriers, it has been assumed that the temperature is low enough for the Fermi--Dirac distribution to be approximated by the Boltzmann distribution (see Appendix~\ref{dielectric}). For laser--excited charge carriers and heavy and light holes, the distribution has been approximated with a step function. This also means that a quasi--equilibrium is assumed to form in the laser--excited semiconductor. Expressions for $n_i$ and $k_{Fi}$ are found in Appendix~\ref{variables}.

If we assume that $\beta_i^2 q^2\ll \w^2+i\gamma\w$, then the expression in Eq.~(\ref{eq:chi2}) can be rewritten by using the fact that it resembles a geometric series to first order. Together with Eq.~(\ref{eq:ep_long4}), we then find that the longitudinal dielectric function is given by
\begin{equation}
\label{eq:ep_long5}
\ep_L(q,\w)=1-\frac{\w_a^2}{\w^2+i\gamma\w-\beta_a^2 q^2}-\frac{\w_b^2}{\w^2+i\gamma\w-\beta_b^2 q^2} ,
\end{equation}
which is almost identical to Eq.~(\ref{eq:ep_long2}) from previous section. The main difference is the presence of $\ep_{\infty}$ in Eq.~(\ref{eq:ep_long2}) which contains the interband transitions. This parameter is simply added ``by hand'', and the value can often be found as a constant in data books. The second discrepancy is the damping constant $\gamma$ which in Eq.~(\ref{eq:ep_long2}) is different for the two charge carriers. Since the charge carriers are expected to have different mobilities $\mu_a$ and $\mu_b$, and the damping constants are related to the mobilities by\cite{lyden64}
\begin{equation}
\label{eq:gamma2}
\gamma_i=\frac{e}{m_{i,{\rm cond}}^*\mu_i} ,
\end{equation}
we will allow $\gamma_i$ to assume different values for the two charge carriers. Note, that the effective mass entering Eq.~(\ref{eq:gamma2}) is the \emph{conductivity effective mass}, while $m_i^*$ used in Eqs.~(\ref{eq:plasma})-(\ref{eq:beta2}) is the \emph{density-of-states effective mass}.

The parameters $\w_i$, $\beta_i$ and $\gamma_i$ will in general be different for the two fluids, but there are situations where they coincide. An intrinsic semiconductor with identical effective masses and mobilities of electrons and holes would according to Eqs.~(\ref{eq:plasma}) and (\ref{eq:beta1}) have the same plasma frequency, $\beta$ and $\gamma$ for the two fluids. A more typical semiconductor where $m_e^* < m_h^*$ could also be modulated to obtain $\beta_e=\beta_h$ by \emph{combining} $p$-doping and laser excitation. A larger density of holes would then be used to compensate for the fact that they are heavier than electrons, and obtaining $m_h^*/m_e^*=k_{Fh}/k_{Fe}$ would according to Eq.~(\ref{eq:beta2}) result in identical $\beta$'s [note that $k_{Fi}^3=3\pi^2 n_i$ according to Eq.~(\ref{eq:k_n})].

\section{Bulk and general properties}
\label{bulk}

In this section, we will analyze some of the general properties of the two-fluid model as well as properties related to the infinite medium. The vector wave equations derived here, will also be used in section \ref{mie}.

\subsection{Normal modes}

For the single-fluid HDM, it has been found useful to derive a set of homogeneous equations for the transversal and longitudinal components of the current density\cite{raza11} as originally introduced by Boardman and Paranjape.\cite{boardman77} We will accordingly derive a set of Boardman equations for the two-fluid model. The first step is to apply either the curl or the divergence to Eqs.~(\ref{eq:hydro2}) whereby a set of equations is obtained for either the transversal or the longitudinal fields, respectively. This is shown in Appendix~\ref{matrix} using a compact matrix notation. Secondly, we introduce the following linear relations for both transversal ($T$) and longitudinal ($L$) current densities
\begin{subequations}
\label{eq:j1j2}
\begin{align}
\label{eq:j1j2a}
\J_a^z&=a_1^z \J_1^z+a_2^z \J_2^z ,\\
\label{eq:j1j2b}
\J_b^z&=b_1^z \J_1^z+b_2^z \J_2^z ,
\end{align}
\end{subequations}
where $z=T,L$. Notice that all current densities share the properties $\Curl \J=\Curl \J^T$ and $\Div \J= \Div \J^L$. We now require that $\J_1$ is independent of $\J_2$ which results in 8 equations in total: For both curl and divergence we get two for both $\J_1$ and $\J_2$. The four equations for the longitudinal fields
\begin{subequations}
\begin{align}
\label{eq:boardman4}
&\left[\beta_a^2 \nabla^2+\omega (\omega+i \gamma_a)-\frac{\omega_a^2}{\ep_{\infty}}\left(1+\frac{b_j^L}{a_j^L}\right)\right]\Div \J_j=0 , \\
\label{eq:boardman5}
&\left[\beta_b^2 \nabla^2+\omega (\omega+i \gamma_b)-\frac{\omega_b^2}{\ep_{\infty}}\left(1+\frac{a_j^L}{b_j^L}\right)\right]\Div \J_j=0 , 
\end{align}
\end{subequations}
with $j=1,2$ are the Boardman equations for the divergence. The four equations for the transversal fields
\begin{subequations}
\begin{align}
\label{eq:boardman6}
&\left[ c^2 \nabla^2+\omega^2\ep_{\infty} - \frac{\omega^2\omega_a^2}{\omega(\omega+i \gamma_a)}\left(1+\frac{b_j^T}{a_j^T}\right)\right]\Curl \J_j=0 , \\
\label{eq:boardman7}
&\left[ c^2 \nabla^2+\omega^2\ep_{\infty} - \frac{\omega^2\omega_b^2}{\omega(\omega+i \gamma_b)}\left(1+\frac{a_j^T}{b_j^T}\right)\right]\Curl \J_j=0 , 
\end{align}
\end{subequations}
are the Boardman equations for the curl. The Boardman equations are useful tools when finding the current densities and the electrical fields, and below we use the Boardman equations for the divergence to find the dispersion relations for the longitudinal fields.

\subsection{Vector wave equation}

When solving Maxwell's equations for any geometry, such as the spherically symmetric systems considered in section~\ref{mie}, a suitable starting point is the vector wave equation. Therefore we will now derive the vector wave equation for both the transversal and the longitudinal electrical fields and simultaneously find the dispersion relations. 

Considering purely transversal fields, Eqs.~(\ref{eq:hydro2}) become
\begin{align}
\label{eq:hydroi_trans}
&\J_i^T=\frac{i\w\ep_0\w_i^2}{\w^2+i\gamma_i\w}\E^T \qquad i=a,b, \\
\label{eq:wave2_trans}
&\nabla^2\E^T+\frac{\w^2}{c^2}\ep_{\infty}\E^T=-i\mu_0\w\left(\J_a^T+\J_b^T\right),
\end{align}
which can be combined directly into the vector wave equation for the transversal field
\begin{equation}
\label{eq:transwave}
\nabla^2\E^T+k_T^2\E^T=0 ,
\end{equation}
where the transversal wavenumber is given by
\begin{equation}
\label{eq:transk}
k_T^2=\frac{\w^2}{c^2}\left(\ep_{\infty}-\frac{\w_a^2}{\w^2+i \gamma_a \w}-\frac{\w_b^2}{\w^2+i \gamma_b \w}\right) .
\end{equation}
Notice that this is consistent with the expression for $\ep_T$ in section~\ref{twofluid}, but differs from that by being valid for any geometry (and not just for the infinite case).

Deriving the vector wave equation for the longitudinal field requires a slightly different procedure. Turning to the Boardman equations~(\ref{eq:boardman4}) and (\ref{eq:boardman5}) with $j=1$, we notice that they both have the form
\begin{equation}
\label{eq:boardman8}
\left(\nabla^2+k^2\right)\nabla\cdot \J_1=0 .
\end{equation}
This also means that the variable $k$ must be the same in both cases
\begin{align}
\label{eq:longk1}
k^2=&\frac{\omega (\omega+i \gamma_a)}{\beta_a^2}-\frac{\omega_a^2}{\beta_a^2\ep_{\infty}}\left(1+\frac{b_1^L}{a_1^L}\right)\nonumber \\
&= \frac{\omega (\omega+i \gamma_b)}{\beta_b^2}-\frac{\omega_b^2}{\beta_b^2\ep_{\infty}}\left(1+\frac{a_1^L}{b_1^L}\right) .
\end{align}
From this we find an expression for the ratio $b_1^L/a_1^L$ which we will call $\alpha_1^L$
\begin{align}
\label{eq:alpha1}
&\frac{b_1^L}{a_1^L}=\alpha_1^L=\frac{\beta_a^2 \ep_{\infty}}{\w_a^2}\frac{1}{2}\Biggl( k_a^2-k_b^2  \nonumber \\
&\qquad\mp\sqrt{\left(k_a^2-k_b^2\right)^2+\frac{4\w_a^2\w_b^2}{\beta_a^2\beta_b^2\ep_{\infty}^2}}\Biggr)\nonumber \\
&k_i^2=\left(\w(\w+i\gamma_i) -\frac{\w_i^2}{\ep_{\infty}}\right)\frac{1}{\beta_i^2} \qquad i=a,b .
\end{align}
Now, the same procedure can be carried out for Eqs.~(\ref{eq:boardman4}) and (\ref{eq:boardman5}) with $j=2$, and this gives us instead $\alpha_2^L=b_2^L/a_2^L$. However, the expression for $\alpha_2^L$ is exactly the same as the one for $\alpha_1^L$ because the Boardman equations for the divergence of $\J_1$ and $\J_2$ are the same. Although this seems strange, it is in fact exactly what we would expect: since we have put no restraints on $\Div \J_1$ and $\Div \J_2$ (or equivalently $\J_1^L$ and $\J_2^L$), they each have to contain both solutions [`$+$' and `$-$' in Eq.~(\ref{eq:alpha1})]. We can therefore chose $\alpha_1^L$ as the `$-$' solution and $\alpha_2^L$ as the `$+$' solution (and this will be done henceforth).

We can also obtain an expression for $k^2$ by inserting either $\alpha_1^L$ or  $\alpha_2^L$ back into Eq.~(\ref{eq:longk1}). The result is two different wavenumbers belonging to $\J_1^L$ and $\J_2^L$ respectively
\begin{equation}
\label{eq:longk}
k_{L,\substack{1 \\ 2}}^2=\frac{1}{2}\Biggl( k_a^2+k_b^2 \pm\sqrt{\left(k_a^2-k_b^2\right)^2+\frac{4\w_a^2\w_b^2}{\beta_a^2\beta_b^2\ep_{\infty}^2}}\Biggr) .
\end{equation}
Here $k$ has been given the subscript `$L$', because it turns out that this is in fact the longitudinal wavenumber. This can be seen by taking the divergence of Ampere's Law and defining the longitudinal fields $\E_1^L$ and $\E_2^L$ where $\E_j^L \propto \J_j^L$. Introducing this into Eq.~(\ref{eq:boardman8}) we find
\begin{equation}
\label{eq:longwave}
\nabla^2\E_j^L+ k_{L,j}^2\E_j^L=0 \qquad j=1,2,
\end{equation}
which is the sought vector wave equation for the longitudinal fields.

\subsection{Dispersion for infinite medium}

\begin{figure}
    \includegraphics[width=\columnwidth]{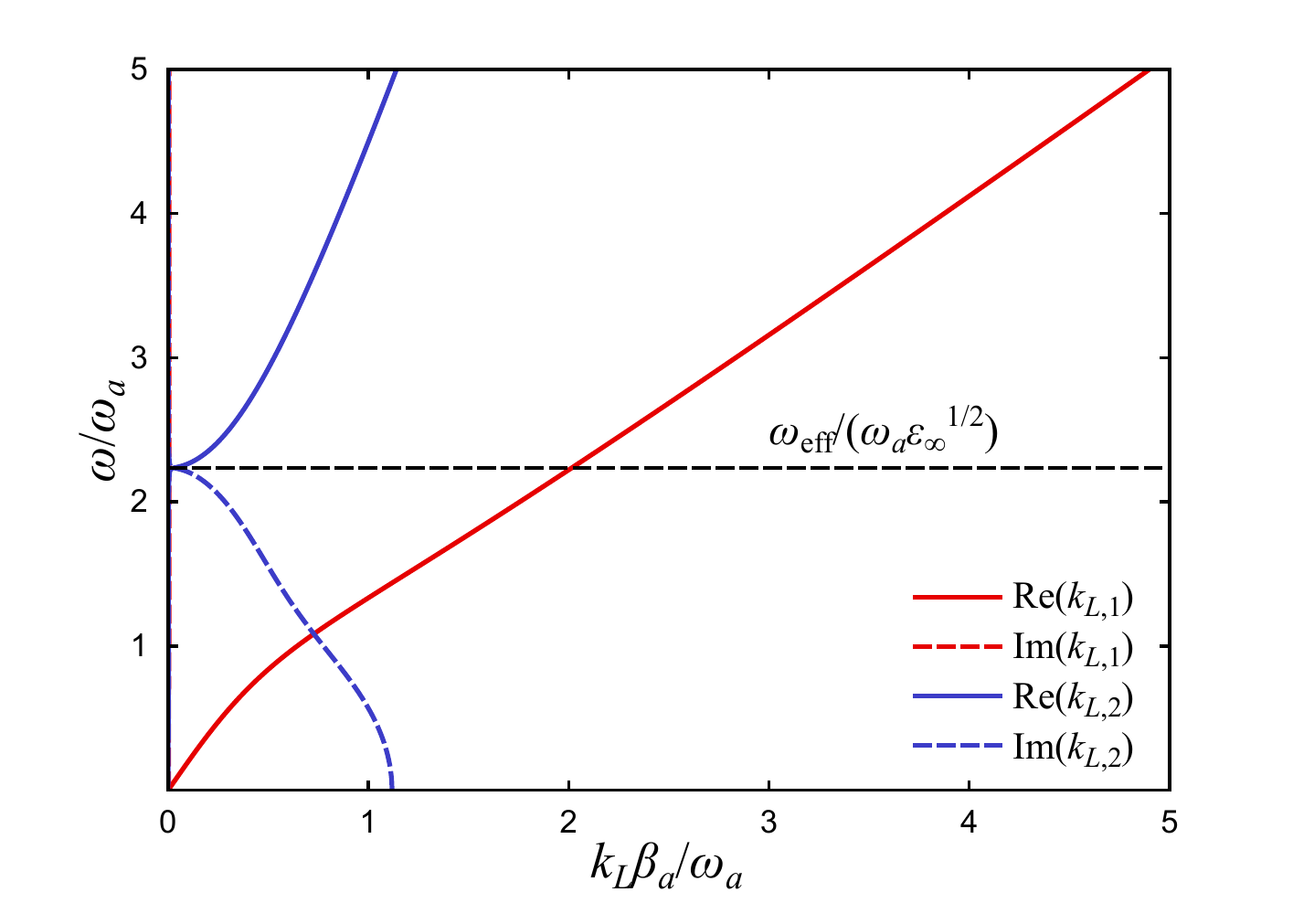}
\caption{\label{fig:dispersion}(Color online) The dispersion relations of the optical mode (blue lines) and the acoustic mode (red lines). The full and dashed lines show the real and imaginary components, respectively, of $k_{L,j}$. The parameters are $\w_b/\w_a=2$, $\beta_b/\beta_a=4$, $\gamma_a=\gamma_b=0.01 \w_a$ and $\ep_{\infty}=1$. Notice that the optical mode has finite imaginary component (i.e. is damped) below $\w_{\rm eff}/\ep_{\infty}^{1/2}$. Due to the small damping constants, the red, dashed line lies almost exactly on top of the $y$-axis.}
\end{figure}

With Eq.~(\ref{eq:longk}) we are now in position to plot the dispersion relations for the longitudinal modes $k_{L,j}(\w)$ for an infinite medium. In Fig.~\ref{fig:dispersion} we show $k_{L,j}(\w)\beta_a/\w_a$ as a function of $\w/\w_a$, and we notice that the two modes ($j=1,2$) have very different appearances. The mode $k_{L,1}(\w)$ follows almost a straight line, while $k_{L,2}(\w)$ is real-valued above a line given by $\w_{\rm eff}/\ep_{\infty}^{1/2}$ with $\w_{\rm eff}^2=\w_a^2+\w_b^2$ and imaginary (damped) below it. Because the $j=2$ mode has non-zero $\w$ for $k_{L,2}\approx 0$, it can be excited by electromagnetic radiation, and for that reason it is denoted the optical mode. The $j=1$ mode is denoted the acoustic mode, and unless methods for momentum matching are applied, it cannot be excited by electromagnetic radiation. The appearance of an optical and an acoustic branch in systems with two different kinds of charge carriers has been observed before in RPA models for infinite media.\cite{pines56,scott93,bonitz00} Here we have found the formation of an optical and an acoustic mode in a two-fluid hydrodynamic model for an infinite medium, something that was briefly touched upon by Schaefer and von~Baltz.\cite{schaefer87} In section~\ref{results} we will analyze both modes in finite systems.

The graphical presentation in Fig.~\ref{fig:dispersion} can be supported by making approximations to Eq.~(\ref{eq:longk}). By isolating the frequency such that we obtain $\w_j(k)$, and taking the limit $k\to 0$, we get the following expressions (ignoring loss)
\begin{subequations}
\label{eq:acoustic_optical}
\begin{align}
\label{eq:acoustic}
\w_1^2(k)&\approx \frac{\w_b^2\beta_a^2+\w_a^2\beta_b^2}{\w_{\rm eff}^2} k^2 , \\
\label{eq:optical}
\w_2^2(k)&\approx \frac{\w_{\rm eff}^2}{\ep_{\infty}}+\frac{\w_a^2\beta_a^2+\w_b^2\beta_b^2}{\w_{\rm eff}^2} k^2 . 
\end{align}
\end{subequations}
Here it is clear that the acoustic mode ($\w_1$) has a linear dependence on $k$, while the optical mode ($\w_2$) mainly is imaginary below the line $\w_{\rm eff}/\ep_{\infty}^{1/2}$.  

In Fig.~\ref{fig:dispersion} we also see that $\mathrm{Im}(k_{L,2})$ is cut off at $\mathrm{Im}(k_{L,2})\beta_a/\w_a\approx 1.1$ for $\w=0$. More generally the cut-off value is $\mathrm{Im}(k_{L,2}) =(\w_a^2/\beta_a^2+\w_b^2/\beta_b^2)^{1/2}/\ep_{\infty}^{1/2}$ as follows from Eq.~(\ref{eq:longk}). This is no unique property of the two-fluid model, and the single-fluid HDM has a similar cut-off at $\w_p/(\beta\ep_{\infty}^{1/2})$ (see the expression for $k_L$ below Eq.~(\ref{eq:mie_delta2}) in Sec.~ \ref{mie}).

That the model contains two longitudinal modes follows directly from the fact that it includes two different kinds of charge carriers. It can be compared with the single-fluid HDM that only has one longitudinal mode. This is an optical mode, i.e. damped below a certain frequency, and for this reason, no longitudinal excitations are expected in this low-frequency region.\cite{raza11} The two-fluid model, on the other hand, also has an acoustic mode which in principle could give rise to excitations below $\w_{\rm eff}/\ep_{\infty}^{1/2}$. In Sec.~\ref{results} we will consider the optical properties of spherical particles. There we will see that peaks indeed emerge in the spectrum below the dipole LSP as a direct consequence of the acoustic mode. In that section we will also show that at higher frequencies, the two fluids will decouple, and the optical response then will resemble that of two independent charge carrier species.

\section{Extended Mie theory}
\label{mie}

We wish analyze the two-fluid model for finite systems, and in this paper we will focus on spherically symmetric systems. Maxwell's equations were originally solved for transversal waves in spherical geometry by Mie,\cite{mie08} and Ruppin later found a solution including longitudinal waves\cite{ruppin73} which has been used together with the HDM for spherical metal particles.\cite{david11,raza13,christensen14} The addition of a second longitudinal wave, however, results in a different system of equations, and here we will derive the Mie coefficients for the two-fluid model.

In spherical geometry, the general solutions to the transversal wave equation [Eq.~(\ref{eq:transwave})] are $\mathbf{m}_{\substack{e\\o}ml}$ and $\mathbf{n}_{\substack{e\\o}ml}$, and the solutions to the longitudinal wave equation [Eq.~(\ref{eq:longwave})] are $\mathbf{l}_{\substack{e\\o}ml}$.\cite{stratton} Here `$e$' and `$o$' are short-hand notation for even and odd, and $m$ and $l$ are integers for which $m\leq l$ holds. We now consider a typical experimental scenario where a plane wave $\E_i$ is incident on a spherical particle which results in a wave scattered (or reflected) from the particle $\E_r$ and a wave transmitted into the particle $\E_t$. Because the functions $\mathbf{m}_{\substack{e\\o}ml}(k_T,\rr)$, $\mathbf{n}_{\substack{e\\o}ml}(k_T,\rr)$ and $\mathbf{l}_{\substack{e\\o}ml}(k_{L,j},\rr)$ form a complete basis, any wave can be written as a linear combination of these. For an $x$-polarized plane wave propagating in the $z$-direction, it can be shown that the linear combination only uses functions of the forms $\mathbf{m}_{o1l}$, $\mathbf{n}_{e1l}$ and $\mathbf{l}_{e1l}$.\cite{stratton} Furthermore, we will assume that the exterior medium is purely dielectric which means that the incident and reflected fields can be written in terms of $\mathbf{m}_{o1l}$ and $\mathbf{n}_{e1l}$ alone. This means that
\begin{align}
\label{eq:Ei}
&\qquad\E_i(\rr,t)= \\
& E_0 e^{-i\w t}\sum_{l=1}i^l \frac{2 l+1}{l (l+1)} \left(\mathbf{m}_{o1l}^{(1)}(k_D,\rr)-i \mathbf{n}_{e1l}^{(1)}(k_D,\rr)\right) , \nonumber\\
\label{eq:Er}
&\qquad\E_r(\rr,t)= \\
& E_0 e^{-i\w t}\sum_{l=1}i^l \frac{2 l+1}{l (l+1)} \left(a_l^r \mathbf{m}_{o1l}^{(3)}(k_D,\rr)-i b_l^r \mathbf{n}_{e1l}^{(3)}(k_D,\rr)\right) , \nonumber
\end{align}
where $k_D=\ep_D^{1/2}\w/c$ and $\ep_D$ is the permittivity of the surrounding dielectric. The superscripts `1' and `3' indicate that the contained spherical Bessel functions are Bessel functions of the first kind ($j_l$) and Hankel functions of first kind ($h_l^{(1)}$), respectively. The expansion coefficients $a_l^r$ and $b_l^r$ in the reflected field are known as the \emph{Mie coefficients}, and the primary goal in this section is to obtain expressions for these.

The transmitted field (i.e. inside the sphere) contains, in addition to the transversal fields, two different longitudinal fields
\begin{align}
\label{eq:Et}
&\E_t(\rr,t)=E_0 e^{-i\w t}\sum_{l=1}i^l \frac{2 l+1}{l (l+1)} \left(a_l^t \mathbf{m}_{o1l}^{(1)}(k_T,\rr)\right.  \\
&\left. -i b_l^t \mathbf{n}_{e1l}^{(1)}(k_T,\rr)+c_{1l}^t \mathbf{l}_{e1l}^{(1)}(k_{L,1},\rr)+c_{2l}^t \mathbf{l}_{e1l}^{(1)}(k_{L,2},\rr)\right) \nonumber,
\end{align} 
where $k_T$ and $k_{L,j}$ are given by (\ref{eq:transk}) and (\ref{eq:longk}), respectively. 

To find the Mie coefficients, a set of suitable boundary conditions (BC) must be provided. By requiring that the fields satisfy Maxwell's Equations and are finite at boundaries, it is found that the parallel components of the electrical and the magnetic fields are continuous, i.e. $\Delta\E_{\parallel}=\mathbf{0}$ and $\Delta\B_{\parallel}=\mathbf{0}$. While these Maxwell BC are sufficient in the local-response solution, additional BC are needed in the two-fluid model. A similar problem was encountered in the HDM where it was found that one additional BC was needed. A physically meaningful BC that is widely used in the HDM is $\J_{\perp}=\mathbf{0}$, which implies that the charge carriers cannot leave surface.\cite{raza15} The two-fluid model requires \emph{two} additional BC, and here we will use the conditions $\J_{b,\perp}=\mathbf{0}$ and $\J_{a,\perp}=\mathbf{0}$. (or equivalently $\J_{1,\perp}=\mathbf{0}$ and $\J_{2,\perp}=\mathbf{0}$). 

Given these BC, we obtain the system of linear equations presented in Appendix~\ref{linear} from which $a_l^{r,t}$, $b_l^{r,t}$ and $c_{jl}^{t}$ can be found. The $a_l^r$ and $b_l^r$ coefficients, which are of primary interest, are given by
\begin{subequations}
\label{eq:mie}
\begin{align}
\label{eq:mie_a}
&a_l^r=\frac{-j_l(x_D)\lbrack x_T j_l(x_T)\rbrack^\prime
             +j_l(x_T)\lbrack x_D j_l(x_D)\rbrack^\prime}
      {h^{(1)}_l(x_D)\lbrack x_T j_l(x_T)      \rbrack^\prime
      -j_l(x_T)      \lbrack x_D h^{(1)}_l(x_D)\rbrack^\prime} , \\
\label{eq:mie_b}
&b_l^r= \nonumber \\
&\frac{-\ep_D j_l(x_D)\left(\Delta_l\!+\!\lbrack x_T j_l(x_T)\rbrack^\prime \right)
                            \!+\!\ep_T j_l(x_T)\lbrack x_D j_l(x_D)\rbrack^\prime}
       {\ep_D h_l^{(1)}(x_D)\left(\Delta_l\!+\!\lbrack x_T j_l(x_T)\rbrack^\prime \right)
                      \!-\!\ep_T j_l(x_T)\lbrack x_D h_l^{(1)}(x_D)\rbrack^\prime} , 
\end{align}
where $x_D=R k_D$ and $x_T=R k_T$. The differentiation (denoted with the prime) is with respect to the argument. The parameter $\Delta_l$ is given by
\begin{equation}
\label{eq:mie_delta}
\Delta_l=\frac{j_l(x_T) l(l+1)}{A}\left(\frac{j_l(x_1) C_2}{x_1 j_l'(x_1)}-\frac{j_l(x_2) C_1}{x_2 j_l'(x_2)}\right) , 
\end{equation}
where $x_j=R k_{L,j}$ and
\begin{align}
\label{eq:mie_C}
C_j=&\frac{\w_a^2\ep_{\infty} k_{L,j}^2}{\beta_a^2\left(1+\frac{1}{\alpha_j^L}\right)}
-\frac{\w_b^2\ep_{\infty} k_{L,j}^2}{\beta_b^2\left(1+\alpha_j^L\right)} , \\
\label{eq:mie_A}
A=&\frac{\left(\w^2+i\gamma_a\w\right)\left(\w^2+i\gamma_b\w\right)(\alpha_1^L-\alpha_2^L)}{\beta_a^2\beta_b^2(1+\alpha_1^L)(1+\alpha_2^L)} ,
\end{align}
\end{subequations}
and $\alpha_j$ is defined in Eq.~(\ref{eq:alpha1}). The coefficients $a_l^r$ are related to oscillations of the magnetic type, and the expression is identical to the one found in the classical, local derivation.\cite{stratton} The coefficients $b_l^r$ are related to oscillations of the electrical type, and the expression is different from the local result \emph{unless} the nonlocal parameter $\Delta_l$ is set to zero. It should also be mentioned that the formula for $b_l^r$ is identical to the one found for the single-fluid HDM,\cite{ruppin73,david11} except that there $\Delta_l$ is given by\cite{david11}
\begin{equation}
\label{eq:mie_delta2}
\Delta_l=\frac{j_l(x_T) j_l(x_L) l(l+1)}{x_L j_l^\prime(x_L)}\left(\frac{\ep_T}{\ep_{\infty}}-1\right) ,
\end{equation}
where $x_T=R\ep_T^{1/2}\w/c$ and $\ep_T$ is given by Eq.~(\ref{eq:ep_trans3}). Also defined is the dimensionless parameter $x_L=R k_L$ where $k_L=(\w^2+i\gamma\w-\w_p^2/\ep_{\infty})^{1/2}/\beta$.

When the expressions for $a_l^r$ and $b_l^r$ have been found, the extinction cross-section is easily calculated with\cite{stratton}
\begin{equation}
\label{eq:sigma}
\sigma_{\rm ext}=-\frac{2 \pi}{k_D^2}\sum_{l=1}(2 l+1)\mathrm{Re}(a_l^r+b_l^r) .
\end{equation}
In the next section, Eq.~(\ref{eq:sigma}) will be used to find the extinction spectra of nanospheres in the two-fluid model.

\section{Numerical results}
\label{results}

In this section, we will present some numerical simulations of the optical properties of both realistic and artificial materials containing two-fluid systems. 

\subsection{Features in the extinction spectrum}

\begin{figure}
    \includegraphics[width=\columnwidth]{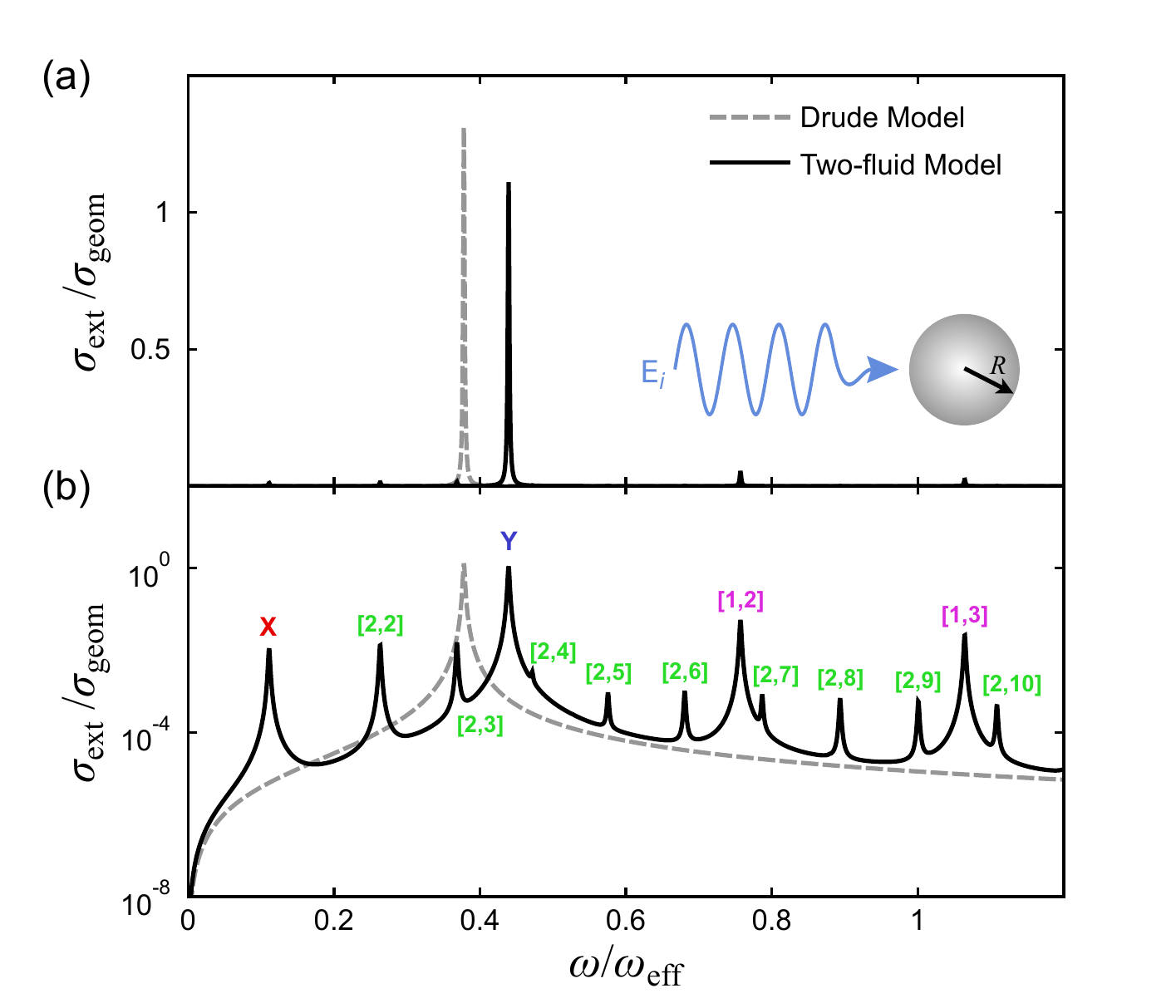}
\caption{\label{fig:extinction}(Color online) (a) The extinction spectrum for Semiconductor A (see parameters in the main text) with $R=10\mathrm{\,nm}$ and $\ep_D=1$. The spectrum has been normalized with $\sigma_{\rm geom}=\pi R^2$. The dashed line is the local Drude model, and the full line is the two-fluid model. (b) The same spectrum plotted with logarithmic $y$-axis. The bulk plasmon peaks are labeled with $[j,n]$, while the first acoustic peak and the LSP peak are indicated with `X' and `Y', respectively.}
\end{figure}

First we will analyze the artificial material `Semiconductor A' with the parameters $\w_a=3.6\times 10^{14}\un{s^{-1}}$, $\w_b=1.8\times 10^{14}\un{s^{-1}}$, $\gamma_a=\gamma_b=1.0\times 10^{12}\un{s^{-1}}$, $\beta_a=4.3\times 10^{5}\un{m\, s^{-1}}$, $\beta_b=1.4\times 10^{5}\un{m\, s^{-1}}$ and $\ep_{\infty}=5$. As we will see later, these parameters are comparable to those of a realistic semiconductor with the exception of the damping constants which have been set low to make the characteristic features of the spectrum clear. We will now consider a spherical particle of this material with $R=10\mathrm{\,nm}$ surrounded by vacuum ($\ep_D=1$). Equation (\ref{eq:sigma}) then gives us the extinction cross section which is shown with the solid line in Fig.~\ref{fig:extinction}(a) as a function of the relative frequency $\w/\w_{\rm eff}$ where $\w_{\rm eff}^2=\w_a^2+\w_b^2$. The spectrum has been normalized with $\sigma_{\rm geom}=\pi R^2$.

The large peak situated around $\w/\w_{\rm eff}=0.4$ can be recognized as the dipole LSP resonance, $\w_{\rm LSP}$, which is also present in the classical local result. However, the peak is shifted to higher frequencies in the two-fluid model as can be seen in the figure by comparing with the local Drude model shown with a dashed line. The local result was found by setting $\Delta_l=0$ in Eq.~(\ref{eq:mie_b}). This blueshift is a well-known nonlocal effect that is also observed in the single-fluid HDM for both metals\cite{raza13b,christensen14} and semiconductors.\cite{maack17} There it is found that the blueshift increases as the particle radius is reduced.

In Fig.~\ref{fig:extinction}(a) we also see small peaks that appear to be present only in the nonlocal model. To investigate this further, the extinction spectrum is shown again in Fig.~\ref{fig:extinction}(b) in a semilogarithmic plot. Now the peaks have become more visible, and several even smaller peaks have appeared. Apart from the LSP resonance, none of these peaks are present in the local solution and, as we will show later, several are not present in the single-fluid HDM either. 

To understand the nature of these resonances, we will consider wavelengths much larger than the particle whereby all the Mie coefficients in Eq.~(\ref{eq:mie_b}) except $b_1^r$ are reduced to zero (see Ref.~\onlinecite{stratton}). Now, when looking for frequencies where the expression diverges, we notice that this occurs whenever $j_1'(x_1) j_1'(x_2)$ in the denominator of $\Delta_1$ vanishes. If we consider the high-frequency region, we can introduce the following large-argument approximation for the spherical Bessel functions\cite{stratton}
\begin{equation}
\label{eq:bessel_approx}
j_l(x_j)\approx \frac{1}{x_j}\cos\left(x_j-\frac{l+1}{2}\pi\right) ,
\end{equation}
and we find that the condition $j_1'(x_1) j_1'(x_2)=0$ is approximately fulfilled whenever $x_j =\pi n$ with $j=1,2$ and $n=1,2,\ldots$. The expression for $k_{L,j}$ in Eq.~(\ref{eq:longk}) can also be simplified at high frequency when $k_a,k_b \gg 2\w_a\w_b/\beta_a\beta_b\ep_{\infty}$ (here ignoring loss)
\begin{align}
\label{eq:longk2}
k_{L,j}^2&\approx\frac{1}{2}\left[\left(\w^2 -\frac{\w_a^2}{\ep_{\infty}}\right)\frac{1}{\beta_a^2}+\left(\w^2 -\frac{\w_b^2}{\ep_{\infty}}\right)\frac{1}{\beta_b^2}\right. \nonumber \\
&\qquad\left.\pm\left(\w^2 -\frac{\w_a^2}{\ep_{\infty}}\right)\frac{1}{\beta_a^2}-\left(\w^2 -\frac{\w_b^2}{\ep_{\infty}}\right)\frac{1}{\beta_b^2}\right] .
\end{align}
Combining this with the condition for $x_j$, we get the following expressions for the resonances
\begin{equation}
\label{eq:quasi_peak2}
\w^2 = \left\{
\begin{array}{rl}
\frac{\pi^2 n^2 \beta_a^2}{R^2}-\frac{\w_a^2}{\ep_{\infty}} \qquad (j=1) \\
\frac{\pi^2 n^2 \beta_b^2}{R^2}-\frac{\w_b^2}{\ep_{\infty}} \qquad (j=2)
\end{array} \right. .
\end{equation}
Here we see that the positions of the peaks are given by two arrays that depend on the properties of either the $a$-fluid \emph{or} the $b$-fluid. In other words, the charge carriers behave as two independent fluids for high frequencies. In Fig.~\ref{fig:extinction}(b), the large peaks above $\w_{\rm LSP}$ can be identified as resonances of the $a$-fluid and are found with the $j=1$ expression, while the small peaks are resonances of the $b$-fluid found with the $j=2$ expression (notice that the distances between the peaks are determined by $\beta_a$ and $\beta_b$). The peaks have been labeled with $[j,n]$, and we notice that $n$ does not start at 1 as is natural to expect. It turns out that the $n=1$ peak simply does not exist and is an artifact of the approximations leading to Eq.~(\ref{eq:quasi_peak2}). 

What is particular noteworthy in the spectrum is that resonances are found in the region below the LSP peak which is ``forbidden'' in the HDM. The reason is that the $a$- and $b$-fluids hybridize and form both an optical and an acoustic branch, where the acoustic branch is characterized by a primarily real wavenumber at frequencies below the LSP peak. This gives rise to the peaks below $\w_{\rm LSP}$ for what reason we will call them \emph{acoustic peaks}. The single-fluid HDM, on the other hand, only contains an optical longitudinal branch which means that no bulk plasmon peaks can exist below the LSP peak.\cite{raza11} 

\begin{figure}
    \includegraphics[width=\columnwidth]{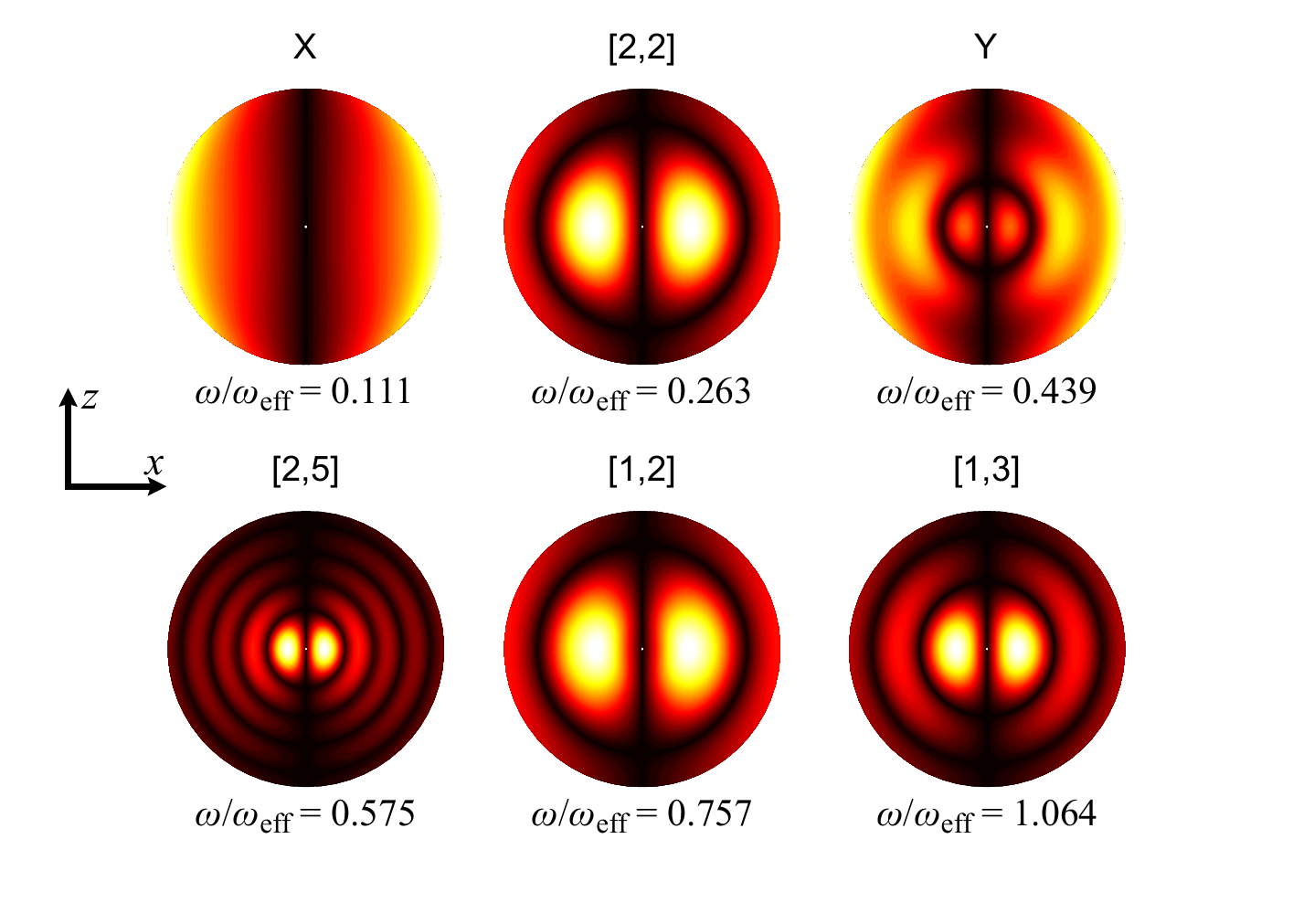}
\caption{\label{fig:charge_dist}(Color online) The charge distribution in the $xz$-plane for Semiconductor A at different frequencies. The damping constants have been set to $\gamma_a=\gamma_b=1.0\times 10^{11}\un{s^{-1}}$ to make the patterns more clear. The incoming wave is directed in the $z$-direction with the electrical field polarized in the $x$-direction. }
\end{figure}

Among the acoustic peaks in Fig.~\ref{fig:extinction}, we find two bulk plasmon peaks labeled with $[2,2]$ and $[2,3]$. However, as a result of the hybridization, these resonances are not purely related to the $b$-fluid, and their positions are therefore only poorly predicted by Eq.~(\ref{eq:quasi_peak2}). Also found below $\w_{\rm LSP}$ is a resonance marked with `X', and it turns out that this is quite different from the bulk plasmons. To see this, the charge distribution inside the sphere is shown in Fig.~\ref{fig:charge_dist} for different frequencies. The contour plots show the distribution in the $xz$-plane when the incoming wave is moving in the $z$-direction, and the electrical field is polarized in the $x$-direction. We here see that the first acoustic peak, marked with `X', is in fact a surface plasmon characterized by a high charge density near the surface. We will discuss this resonance in detail below. The resonance marked with $[2,2]$, on the other hand, is a bulk plasmon with a high charge density near the center, and its distribution is nearly identical to the one marked with $[1,2]$ which is the bulk plasmon of the $a$-fluid of the same order. The peaks marked with $[2,5]$ and $[1,3]$ are bulk plasmons of higher orders for the $b$-fluid and the $a$-fluid, respectively. The charge distribution for the LSP peak is also shown (marked with `Y'), and we see from the contour plot that although it is indeed a surface plasmon, it also displays the pattern of a confined bulk plasmon. The reason is that the LSP resonance hybridizes with the $b$-fluid bulk plasmon marked by $[2,4]$, resulting in a charge distribution with features from both surface and bulk plasmons. Such a hybridization would never take place in the HDM where the surface plasmons always are clearly separated in frequency from the bulk plasmons.

Notice that all the charge distributions are dipole modes, i.e. symmetric along the direction of the $\E$-field, and the same is true for all the visible peaks in Fig.~\ref{fig:extinction}(b). A family of higher-order modes in fact does exist for each peak, but they are too faint to be seen in this spectrum (see Ref.~\onlinecite{christensen14} for an analysis of multipoles in the HDM).

\begin{figure}
    \includegraphics[width=\columnwidth]{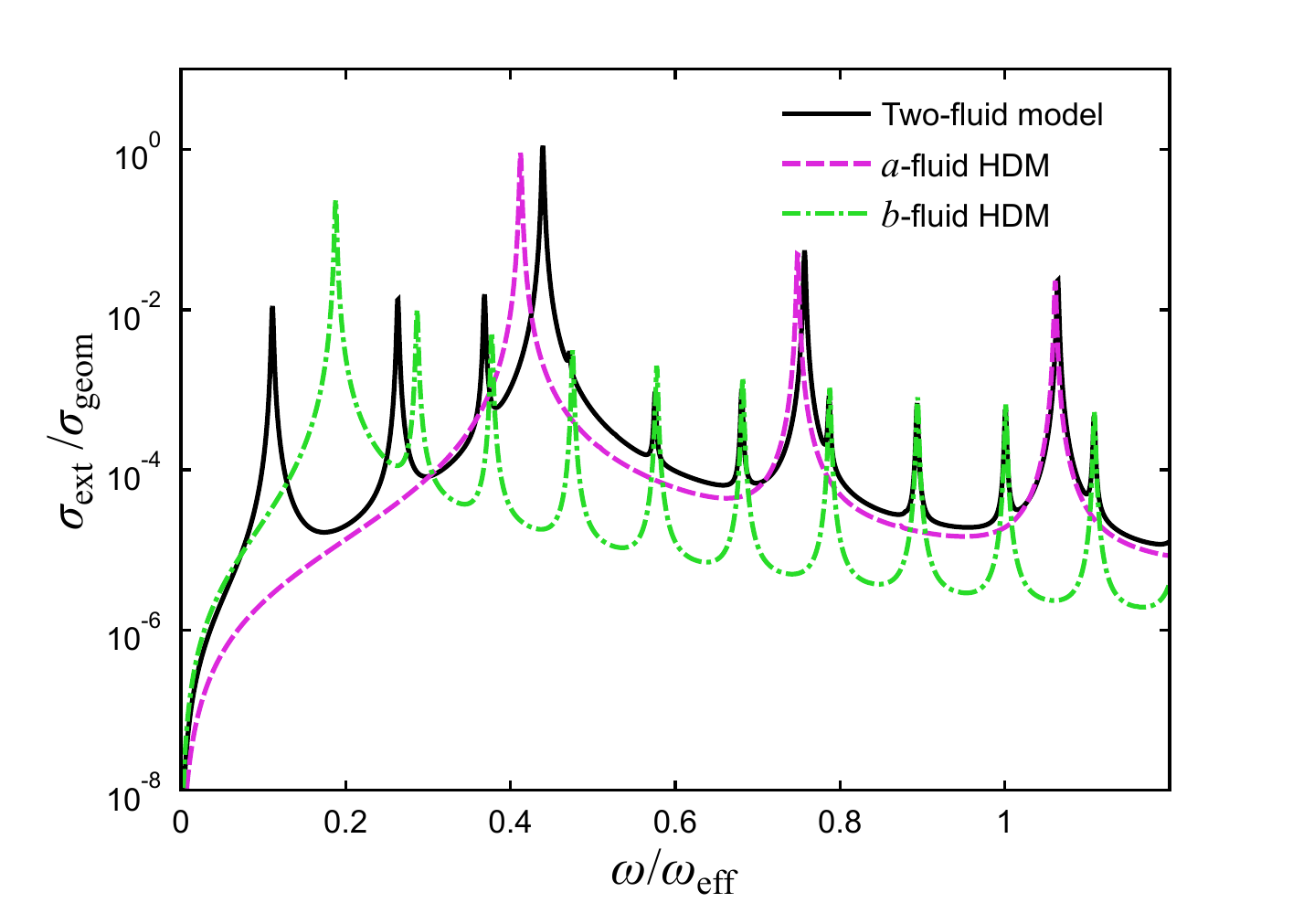}
\caption{\label{fig:single_fluid}(Color online) The spectrum of Semiconductor A (parameters given in the main text) as found by the two-fluid model is shown with the full, black line. The dashed, magenta line and the dash-dotted, green line show the spectra found with the single-fluid HDM when including the $a$-fluid and $b$-fluid, respectively.}
\end{figure}

\subsection{Comparison to the HDM}

It has already been indicated that the two-fluid model is similar to the traditional single-fluid HDM on some points and different on others. To analyze the differences, the extinction spectra for Semiconductor A as calculated by the two different models are shown in Fig.~\ref{fig:single_fluid} for $R=10\mathrm{\,nm}$ and $\ep_D=1$. The extinction cross section has been calculated for the single-fluid HDM by only including one kind of charge carrier and ignoring the other (this was also done in Ref.~\onlinecite{maack17}). In this case, the single-fluid parameters are given by $\w_p=\w_i$, $\beta=\beta_i$ and $\gamma=\gamma_i$, and the nonlocal parameter $\Delta_l$ is found with Eq.~(\ref{eq:mie_delta2}) rather than (\ref{eq:mie_delta}).

When the $a$-fluid is included in the single-fluid HDM, the spectrum with the dashed, magenta line is obtained, and we see that it reproduces the $j=1$ bulk plasmon peaks found in the two-fluid model very well. This is related to the fact that the bulk plasmon peaks in the two-fluid model mainly are determined by the properties of the charge carriers separately, as was indicated in Eq.~(\ref{eq:quasi_peak2}). Additionally, the LSP peak in the single-fluid model is almost at the same position as the one in the two-fluid model.

\begin{figure}
    \includegraphics[width=\columnwidth]{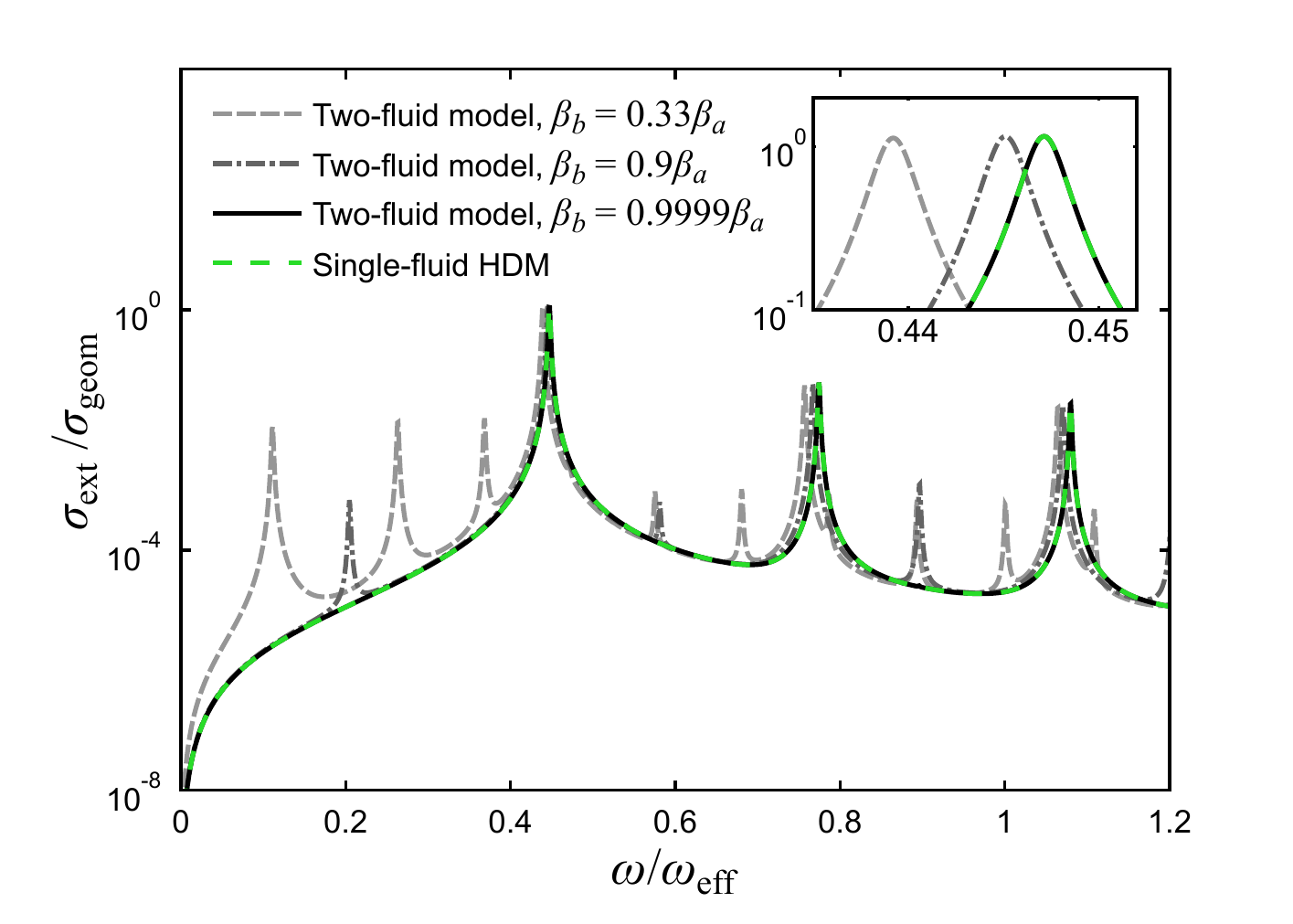}
\caption{\label{fig:same_beta}(Color online) The spectrum of Semiconductor A (parameters given in the main text) as found by the two-fluid model for different values of $\beta_b$: $0.33\beta_a$, $0.9\beta_a$ and $0.9999\beta_a$. When the $\beta$'s approach the same value, the spectrum coincides with the one predicted by the single-fluid HDM (shown with dashed, green). }
\end{figure}

The dash-dotted, green line in the figure shows the extinction cross section for the single-fluid HDM when only the $b$-fluid is included, and we see that it matches well with the $j=2$ bulk plasmon peaks in the two-fluid model. It also reproduces two of the acoustic peaks reasonably well, but is completely off when it comes to the first acoustic peak (marked with `X' in Fig.~\ref{fig:extinction}(b)). This first peak is therefore a feature of the two-fluid model that cannot be reproduced by two independent single-fluid models. 

It was mentioned in section~\ref{twofluid} that the two-fluid model reduces to the single-fluid model if $\beta_a=\beta_b$ and $\gamma_a=\gamma_b$. This is shown in Fig.~\ref{fig:same_beta} where the extinction spectrum for Semiconductor A in the two-fluid model is plotted for increasingly similar $\beta$ values. The green dashed line in the figure shows the single-fluid HDM with $\w_p^2=\w_a^2+\w_b^2$ and $\beta=\beta_a$, and we see that it is exactly on top of the line showing the $\beta_b=0.9999\beta_a$ case. Confirming that the two-fluid model reduces to the single-fluid HDM for $\beta_a=\beta_b$ is also a corroboration of the numerical results. Finally, it is worth mentioning that the local approximation, $\beta_a=\beta_b=0$, is a special case of identical $\beta$'s. This can be understood from the fact that in the Drude model, both current densities are directly proportional to the electrical field, which means that they always can be collected into an effective current density (still assuming that $\gamma_a=\gamma_b$).

\begin{figure}
    \includegraphics[width=\columnwidth]{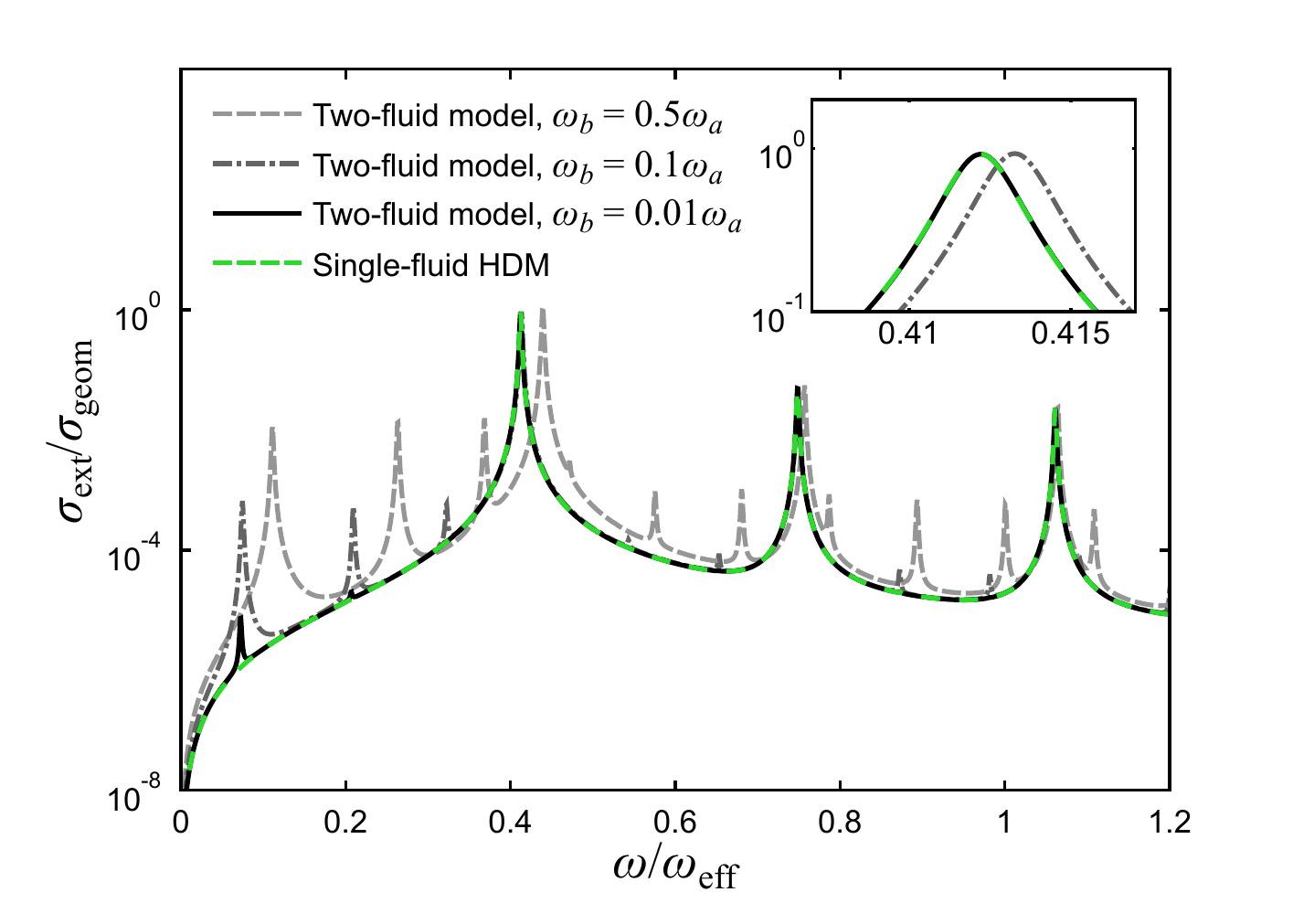}
\caption{\label{fig:low_wph}(Color online) The spectrum of Semiconductor A (parameters given in the main text) as found by the two-fluid model for different values of $\w_b$: $0.5\w_a$, $0.1\w_a$ and $0.01\w_a$. As $\w_b$ diminishes, the spectrum becomes identical to the one obtained by the single-fluid HDM (shown with dashed, green). }
\end{figure}

Apart from the singular situation where the $\beta$'s and $\gamma$'s are identical, the two-fluid model should ideally always be applied to semiconductors where two kinds of charge carriers are present. But as noted in the Introduction, materials where electrons are present as majority carriers can effectively be considered single-fluid systems. The smaller effective mass and larger density of electrons compared to holes will according to Eq.~(\ref{eq:plasma}) result in a much larger plasma frequency. And this in turn causes the electrons to determine the optical properties almost completely, which means that it is sufficient to use the single-fluid HDM. In Fig.~\ref{fig:low_wph}, the spectrum of Semiconductor~A is shown for various values of $\w_b$. We see that for $\w_b=0.01\w_a$, almost all unique features of the two-fluid model are gone, and the spectrum coincides with the one predicted by the single-fluid HDM including only charge carrier $a$. Ratios of $0.01$ between the plasma frequencies are easily obtained in doped semiconductors. If we consider an $n$-doped semiconductor with $n=10^{18}\un{cm^{-3}}$ and an intrinsic carrier concentration of $n_{\rm int}=10^{16}\un{cm^{-3}}$, the fundamental relation\cite{sze} $n_{\rm int}^2=n p$ tells us that the hole concentration will be $p=10^{14}\un{cm^{-3}}$. Accounting for the larger mass of the holes ('$h$') compared to the electrons ('$e$') we indeed obtain $\w_h/\w_e<0.01$. For this reason we propose the two-fluid model for $p$-doped systems and systems where $\w_h/\w_e>0.1$.

\subsection{Indium antimonide and gallium arsenide}

After analyzing the artificial material Semiconductor A, we will now look at more realistic semiconductors. The first material we will consider is intrinsic InSb where the electrons are thermally excited across the bandgap. As seen in Table~\ref{tab:data}, InSb has a very narrow bandgap which gives rise to relatively high charge carrier densities even at room temperature. If we choose $T=300\un{K}$, electrons ($e$) as the $a$-fluid, holes ($h$) as the $b$-fluid and use Eqs.~(\ref{eq:plasma}), (\ref{eq:beta1}) and (\ref{eq:gamma2}) from section~\ref{micro} and Eq.~(\ref{eq:n_therm}) from Appendix~\ref{variables} together with data from Table~\ref{tab:data}, we then find $\w_e=6.09\times 10^{13}\un{s^{-1}}$, $\w_h=1.07\times 10^{13}\un{s^{-1}}$, $\gamma_e=1.99\times 10^{12}\un{s^{-1}}$, $\gamma_h=6.67\times 10^{12}\un{s^{-1}}$, $\beta_e=1.13\times 10^{6}\un{m\, s^{-1}}$ and $\beta_h=1.99\times 10^{5}\un{m\, s^{-1}}$. 

\begin{table}
\caption{Data for GaAs and InSb. The intrinsic charge carrier density is denoted by $n_i$. The masses $m_e^*$ and $m_{hh}^*$ for InSb are taken from Refs.~\onlinecite{stradling70} and \onlinecite{oszwaldowski87}, respectively. For GaAs, $m_e^*$ and $m_{e,{\rm cond}}^*$ (which depends on the doping level $N_a$) are from Ref.~\onlinecite{szmyd90}, and $m_{lh}^*$ and $m_{hh}^*$ are from Ref.~\onlinecite{walton68}. $E_g$ for InSb is taken from Ref.~\onlinecite{rowell88}, and $\mu_e$ and $\mu_h$ for GaAs are from Ref.~\onlinecite{sze67}. The rest of the data are taken from Ref.~\onlinecite{madelung}. Note that for InSb, the value of $m_{e,{\rm cond}}^*$ is assumed to be identical to $m_e^*$, and $m_{lh}^*$ is the $0\un{K}$ value.}
\begin{tabular}{lrrrr}
\hline\hline\\[-5pt]
 & \multicolumn{2}{r}{GaAs ($300\mathrm{\,K}$)} & InSb ($300\mathrm{\,K}$) & InSb ($400\mathrm{\,K}$) \\[5pt]
\hline
$\ep_{\infty}$            & & $ 10.86 $            & $ 15.68 $             & $ 15.68 $           \\
$E_g$ (eV)                & & $ 1.424 $            & $ 0.174 $             & $ 0.146 $           \\
$n_i$ ($\cmm$)         & & $ 2.18\times 10^6 $ & $ 1.34\times 10^{16} $ & $ 5.73\times 10^{16} $ \\
\mc{\mr{$\mu_e$ ($\mob$)}}  & $ 7000 $\footnotemark[1] & \mr{$ 77000 $}   & \mr{$ 48000 $}       \\
\mc{}                       & $ 1100 $\footnotemark[2] &                   &                      \\
\mc{\mr{$\mu_h$ ($\mob$)}}  & $ 400 $\footnotemark[1]  & \mr{$ 850 $}      & \mr{$ 480 $}        \\
\mc{}                       & $ 80 $\footnotemark[2]   &                   &                      \\
$m_e^*/m_0$               & & $ 0.0636 $          & $ 0.0115 $             & $ 0.0100 $          \\
\mr{$m_{e,{\rm cond}}^*/m_0$} & & $ 0.0636 $\footnotemark[1] & \mr{$ 0.0115 $} & \mr{$ 0.0100 $} \\
                          & & $ 0.1014 $\footnotemark[2]     &             &                      \\
$m_{lh}^*/m_0$            & & $ 0.093 $           & $ 0.016 $              & $ 0.016 $           \\
$m_{hh}^*/m_0$            & & $ 0.50 $            & $ 0.37 $               & $ 0.40 $            \\
\hline\hline
\end{tabular}
\footnotetext[1]{$N_a=0\un{cm^{-3}}$}
\footnotetext[2]{$N_a=10^{19}\un{cm^{-3}}$}
\label{tab:data}
\end{table}

\begin{figure}
    \includegraphics[width=\columnwidth]{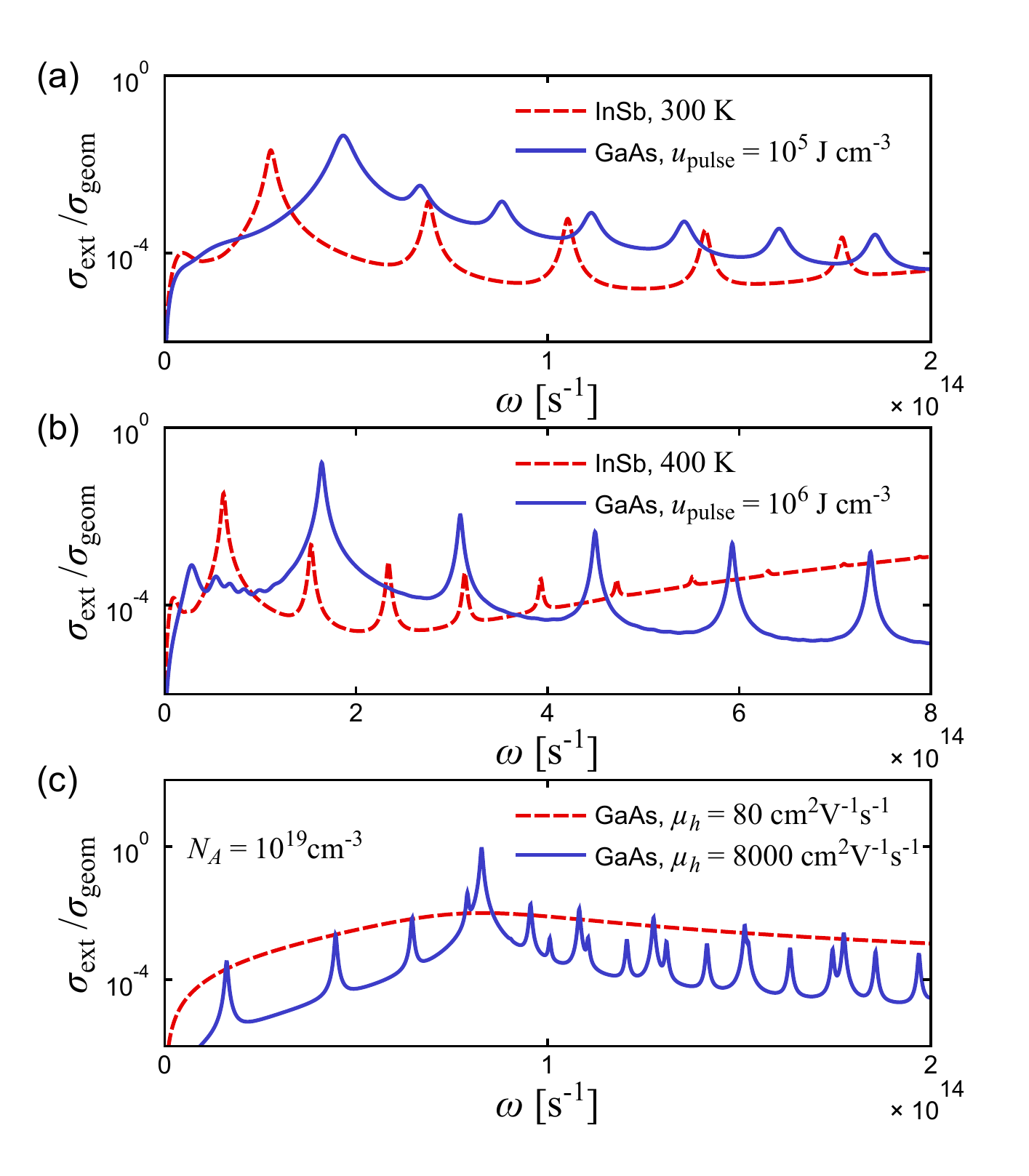}
\caption{\label{fig:insb_gaas}(Color online) Extinction spectra for InSb and GaAs. In all cases is $\ep_D=1$. (a) The dashed, red line and the full, blue line show the spectra for intrinsic InSb at $300\un{K}$ with $R=100\un{nm}$ and laser excited GaAs with $u_{\rm pulse}=10^5\un{J\, cm^{-3}}$ and $R=40\un{nm}$, respectively. (b) The dashed, red line and the full, blue line show the spectra for intrinsic InSb at $400\un{K}$ with $R=60\un{nm}$ and laser excited GaAs with $u_{\rm pulse}=10^6\un{J\, cm^{-3}}$ and $R=15\un{nm}$, respectively. (c) The spectrum for $p$-doped GaAs with $N_A=10^{19}\un{cm^{-3}}$ and $R=30\un{nm}$ is shown with the dashed, red line. For the full, blue line  the mobility of the holes is artificially set a hundred times higher.}
\end{figure}

Considering a very small particle of InSb would give us clearly visible nonlocal effects which are interesting in terms of analyzing the model, but the number of charge carriers, which scale as $R^{-3}$, would also be smaller. And at some point there will be too few charge carriers for them to be considered a plasma, which means that a plasma model no longer is suitable. Therefore we will choose the radius of the InSb particle to be $100\un{nm}$ which results in the number of electrons and holes to be $N_e=N_h=56$. If we then choose the surrounding medium to be vacuum, we find the extinction spectrum shown in Fig.~\ref{fig:insb_gaas}(a) with the dashed, red line. Here we see the LSP peak at $\sim\! 3\times 10^{13}\un{s^{-1}}$ followed by several electron bulk plasmon peaks. The hole plasmon peaks are completely invisible, a result of the size of the particle and the low mobility of the holes. However, one of the acoustic peaks is still visible, which could be interesting in terms of verifying the model. 

The full, blue line in Fig.~\ref{fig:insb_gaas}(a) shows the extinction spectrum of a $40\un{nm}$, intrinsic GaAs particle in vacuum with electrons excited to the conduction band by a laser pulse. The pulse has an energy density of $u_{\rm pulse}=10^5\un{J\, cm^{-3}}$ which results in a number of electrons and holes of $N_e=N_h=114$. Using the equations from section~\ref{micro} and Appendix~\ref{variables} we find $\w_e=1.46\times 10^{14}\un{s^{-1}}$, $\w_h=5.07\times 10^{13}\un{s^{-1}}$, $\gamma_e=3.95\times 10^{12}\un{s^{-1}}$, $\gamma_h=1.16\times 10^{13}\un{s^{-1}}$, $\beta_e=3.29\times 10^{5}\un{m\, s^{-1}}$ and $\beta_h=3.95\times 10^{4}\un{m\, s^{-1}}$. We here recognize the largest peak as the LSP peak followed by a series of electron bulk plasmon peaks, while the hole bulk plasmon peaks are completely suppressed by damping.

It is also interesting to consider a higher temperature for the InSb particle and a stronger laser pulse for the GaAs particle. The dashed, red line in Fig.~\ref{fig:insb_gaas}(b) shows an intrinsic InSb particle with $R=60\un{nm}$ at $400\un{K}$ which results in $N_e=N_h=51$, and the full, blue line shows an intrinsic GaAs particle with $R=15\un{nm}$ and $u_{\rm pulse}=10^6\un{J\, cm^{-3}}$ which results in $N_e=N_h=57$. Here the acoustic peaks, one of the interesting features of the spectra, are more visible.

To analyze the two-fluid model for semiconductors with light and heavy holes, we will consider $p$-doped GaAs with an acceptor concentration of $N_A=10^{19}\un{cm^{-3}}$. According to the equations of section~\ref{micro} and Appendix~\ref{variables}, this results in a concentration of light and heavy holes of $n_{lh}=7.43\times 10^{17}\un{cm^{-3}}$ and $n_{hh}=9.26\times 10^{18}\un{cm^{-3}}$ and the parameters $\w_{lh}=1.59\times 10^{14}\un{s^{-1}}$, $\w_{hh}=2.43\times 10^{14}\un{s^{-1}}$, $\gamma_{lh}=\gamma_{hh}=5.79\times 10^{13}\un{s^{-1}}$, $\beta_{lh}=2.70\times 10^{5}\un{m\, s^{-1}}$ and $\beta_{hh}=1.17\times 10^{5}\un{m\, s^{-1}}$. Here it has been assumed that the light and heavy holes have the same damping constant which is found with $\mu_h$ from Table~\ref{tab:data} and $m_{h,\mathrm{cond}}$ from Eq.~(\ref{eq:m_effective2}). Choosing $R=30\un{nm}$ and $\ep_D=1$ produces the extinction spectrum shown with the dashed, red line in Fig.~\ref{fig:insb_gaas}(c). Here the only visible feature is the LSP peak, while the bulk plasmons are completely damped. For the purpose of analyzing the model, the full, blue line shows the spectrum for the same material, but with the mobility of the holes set a hundred times larger. Now we see the bulk plasmon peaks for both charge carriers as well as the peaks below $\w_{\rm LSP}$. 

Another group of semiconductors that is gaining increasing popularity as plasmonic materials is the transparent conducting oxides (TCO) such as indium tin oxide (ITO), aluminum-doped ZnO (AZO) and indium-doped CdO (In:CdO). ITO was used in Refs.~\onlinecite{garcia11}, \onlinecite{liu17} and \onlinecite{liu18}, and In:CdO was used in Refs.~\onlinecite{yang17} and \onlinecite{deceglia17a}. Apart from the advantages that TCO's share with other semiconductors (such as tunability), they are particularly suitable for the creation of thin films and often allow for heavy doping.\cite{naik13} The most commonly used TCO's, including ITO, AZO and In:CdO, are $n$-type semiconductors\cite{comin14} (ZnO and CdO are even $n$-type semiconductors without intentional doping\cite{chopra83,sachet15}) and as established above, materials with electrons as majority carriers can be modeled with the single-fluid HDM. However, much effort is currently going into the development of $p$-type TCO's,\cite{scanlon12,zhang16,tang17} and it is not unlikely that  TCO's suitable for investigating the two-fluid model will be discovered.

In our model, we have left out some of the mechanisms found in real semiconductors. As mentioned in section~\ref{micro}, interband transitions are ignored, and the effects of them are assumed to be contained in $\ep_{\infty}$. This is a reasonable approximation as long as the energies considered are smaller than the bandgap. Some semiconductors also contain excitons which are caused by the Coulomb interaction between electrons and holes and give rise to energy levels inside the bandgap. However, for doped semiconductors and intrinsic semiconductors with narrow bandgaps, the screening from the high density of charge carriers significantly weakens the binding energy of the excitons.\cite{haug} It is therefore a decent approximation for these materials to leave out excitons. A third kind of excitation especially found in non-elemental semiconductors is optical phonons. These resonances of the lattice may couple to the plasmons if they are in the same frequency range, and this interaction has been studied for both InSb\cite{gaur76,gu00} and GaAs.\cite{olsen69,chen89} It should also be mentioned that for InSb in particular, a charge-carrier depleted region known as the \emph{space-charge layer} may exist close to the surface which would be relevant for the optical properties. This layer has been investigated in several earlier papers,\cite{ritz85,jones95,bell98,adomavicius09} and the question of how it affects features such as the acoustic peak still remains. Finally the two-fluid model, just as the single-fluid HDM, does not account for Landau damping whereby the energy of the plasmons dissipates into single-particle excitations. The excitation of single particles depends on the momentum $q$, and in that sense Landau damping is a size-dependent, nonlocal loss mechanism. Although not considered here, nonlocal damping could be incorporated into the two-fluid model by allowing the $\beta$'s to become complex as it is done for the single-fluid HDM in Refs.~\onlinecite{mortensen14}, \onlinecite{raza15}, and~\onlinecite{deceglia17a}.

\subsection{The acoustic peaks}

One of the defining characteristics of the two-fluid model is the presence of resonances below $\w_{\rm LSP}$, and the experimental observation of these could potentially be used to verify model. Therefore this section will be used to analyze these acoustic peaks, and the focus will be on the first acoustic peak [marked with `X' in Fig.~\ref{fig:extinction}(b)]. 

\begin{figure}
    \includegraphics[width=\columnwidth]{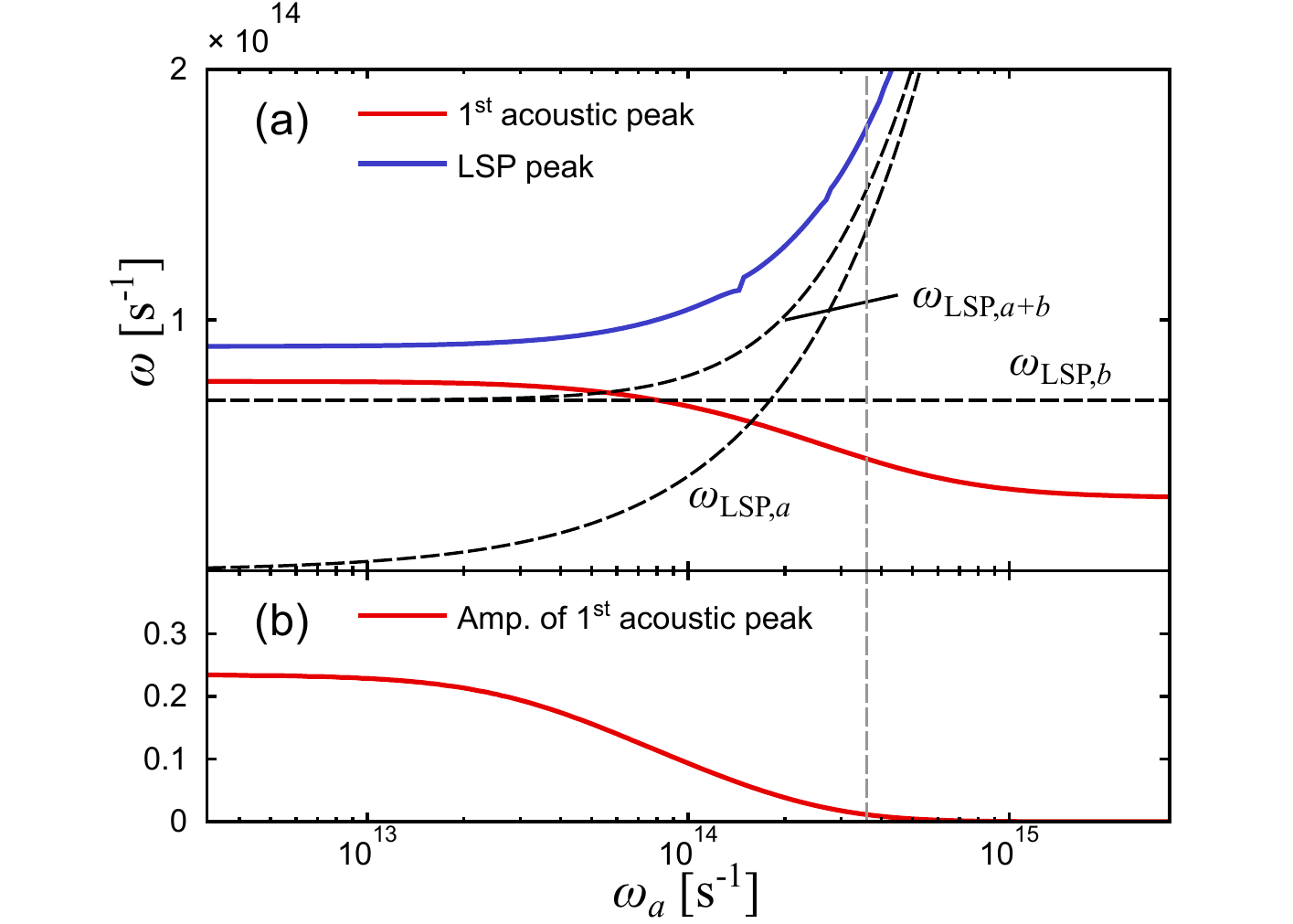}
\caption{\label{fig:acoustic_vs_wp}(Color online) (a) The spectral position of the first acoustic peak as a function of $\w_a$ for Semiconductor A is shown with a red line, while the blue line shows the position of the LSP peak. The vertical, dashed line shows the value $\w_a=3.6\times 10^{14}\un{s^{-1}}$ which was used in previous figures with Semiconductor A. The dent in the blue line around $\w_a\approx 10^{14}\un{s^{-1}}$ is not a numerical artifact, but is caused by bulk plasmon resonances that are superimposed on the LSP peak and thereby make the definition of the peak ambiguous. (b) The amplitude of the first acoustic peak normalized with $\sigma_{\rm geom}$ as a function of $\w_a$.}
\end{figure}

We will start by considering the artificial material Semiconductor A, and in Fig.~\ref{fig:acoustic_vs_wp}(a), the spectral positions of the first acoustic peak and the LSP peak are shown as functions of $\w_a$ with a red and blue line, respectively. Here it is interesting to note that while the LSP peak blueshifts as $\w_a$ increases, the acoustic peak instead moves to lower frequencies. Also shown in the figure with dashed, black lines is the position of LSP peak in the local response approximation as given by $\w_{\mathrm{LSP},a+b}=(\w_a^2+\w_b^2)^{1/2}/(\ep_{\infty}+2\ep_D)^{1/2}$ (including both kinds of charge carriers) and $\w_{\mathrm{LSP},i}=\w_i/(\ep_{\infty}+2\ep_D)^{1/2}$ (including charger carrier $a$ or $b$). The vertical, dashed line marks $\w_a=3.6\times 10^{14}\un{s^{-1}}$ which was used in previous figures with Semiconductor A

Apart from the position of the acoustic peak, the amplitude will also play a role, especially in terms of detecting the resonance. Fig.~\ref{fig:acoustic_vs_wp}(b) shows the amplitude of the first acoustic peak, and we here see that it decreases when $\w_a$ goes up.

\begin{figure}
    \includegraphics[width=\columnwidth]{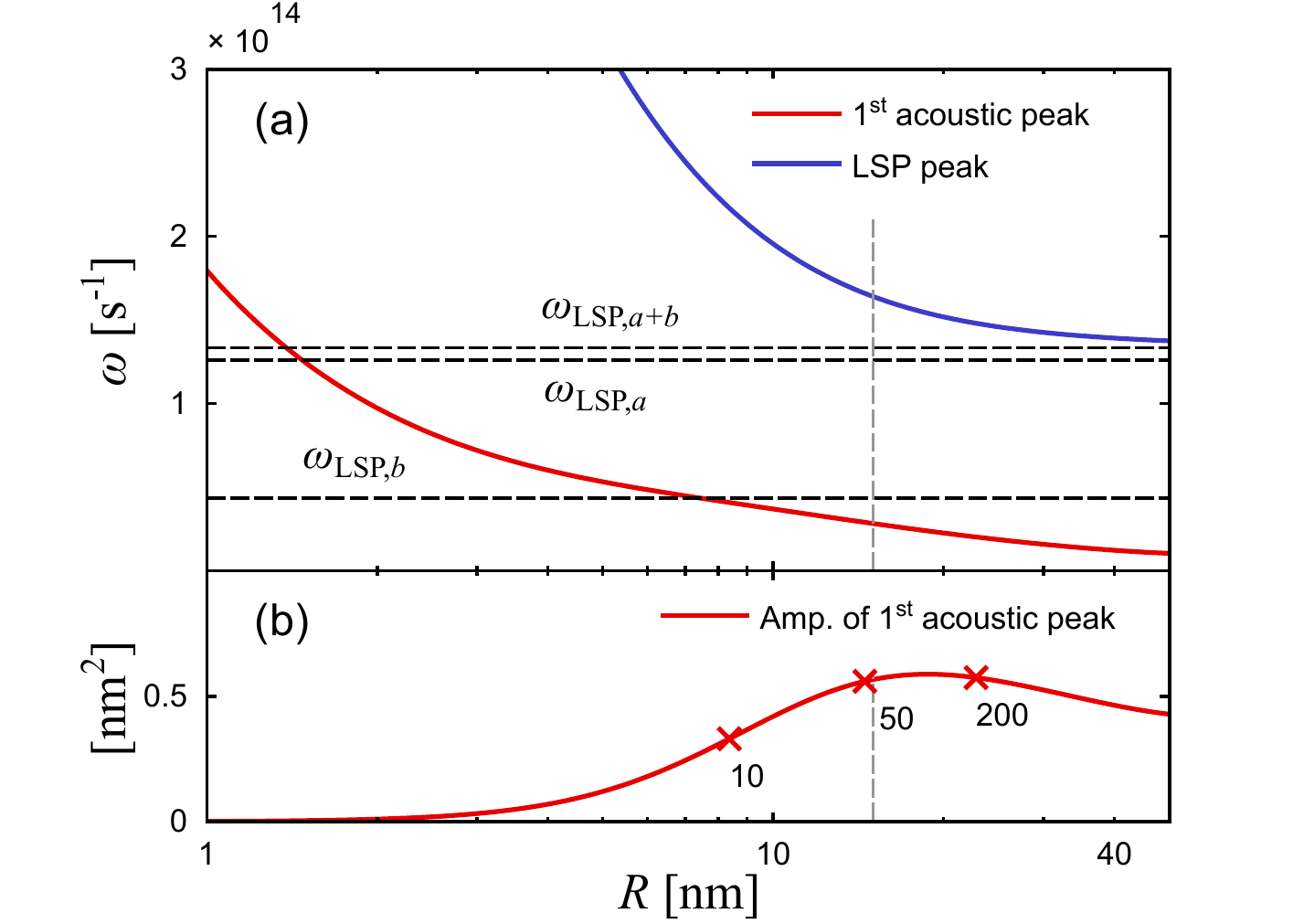}
\caption{\label{fig:acoustic_vs_r}(Color online) (a) The spectral positions of the first acoustic peak and the LSP peak as functions of $R$ for intrinsic GaAs shown with a red and blue line, respectively. GaAs was excited with a laser pulse of $u_{\mathrm{pulse}}=10^6\un{J\, cm^{-3}}$. The vertical, dashed line marks the value $R=15\un{nm}$ which was used in Fig.~\ref{fig:insb_gaas}(b). (b) The amplitude of the first acoustic peak as a function of $R$ (unnormalized). The number of electrons in the particle is indicated for three different sizes.}
\end{figure}

Turning to intrinsic GaAs, Fig.~\ref{fig:acoustic_vs_r}(a) shows the spectral positions of the first acoustic peak and the LSP peak as functions of the radius of the particle. The particle has been excited by a laser pulse of $u_{\mathrm{pulse}}=10^6\un{J\, cm^{-3}}$ and is surrounded by vacuum. Here we see that the position of the acoustic peak, shown with a red line, blueshifts when $R$ is reduced. The LSP peak, shown with a blue line, also blueshifts which is similar to what is found in the HDM for both metals\cite{raza13,christensen14} and semiconductors.\cite{maack17}  

Fig.~\ref{fig:acoustic_vs_r}(b) shows the amplitude of the first acoustic peak, and it is interesting to see that the height of the peak reaches a maximum around $R=20\un{nm}$. The number of electrons in the particle is also given in the figure for three different particle sizes. Note that the extinction cross section in this figure is the absolute value, since normalization with $\pi R^2$ would make the interpretation of the results more difficult.

As the first acoustic peak could be used to verify the model, it is relevant to find the scenario where this resonance is easiest to detect. Fig.~\ref{fig:acoustic_vs_wp} shows the amplitude and position of the peak as functions of $\w_a$ for Semiconductor A, but for a realistic semiconductor it will not be possible to freely vary this parameter. For laser-excited GaAs, Fig.~\ref{fig:acoustic_vs_r}(b) shows, interestingly, that the amplitude of the acoustic peak reaches a maximum for a certain finite radius, and a similar behavior is expected for other materials and geometries. 

The materials investigated here are not all equally well suited to test the model. In the case of $p$-doped GaAs, it was found that none of the features of the two-fluid model are present due to the low mobility of the holes. However, a $p$-doped semiconductor with higher mobility of the holes might still be used to test the two-fluid model. In the case of laser-excited GaAs with $u_{\rm pulse}=10^6\un{J\, cm^{-3}}$, clear acoustic peaks were found, but it must be remembered that the charge carriers will decay over time which will create new experimental opportunities \emph{and} challenges. Finally, intrinsic InSb with thermally excited charge carriers is perhaps the best candidate in terms of testing the model, as the spectrum remains stable over time and is expected to contain the acoustic peaks.

\section{Conclusions}

The hydrodynamic Drude model (HDM), which has successfully described the optical properties of metallic nanostructures, can be adapted to semiconductors by accounting for the fact that several different kinds of charge carriers are present. In this paper, we have presented a two-fluid hydrodynamic model for semiconductors containing electrons and holes or light and heavy holes. We have shown that the two-fluid model is supported by a microscopic theory, and simultaneously we found expressions for the nonlocal parameter $\beta$ for thermally excited charge carriers, laser excited charge carriers and $p$-doped semiconductors with light and heavy holes. 

It was found that the two hydrodynamic fluids hybridize to form an acoustic and an optical branch, both longitudinal, whereas the single-fluid HDM only contains an optical branch. An extended Mie-theory was developed to accommodate the two longitudinal waves, and this theory was subsequently applied to semiconductor nanospheres to find the extinction spectra. We found that in addition to the well-known features of the single-fluid HDM, the two-fluid model displays at least two new optical features: 1) a second set of bulk plasmon resonances and 2) acoustic resonances below the dipole LSP peak, of which the first attains its maximal strength at a finite particle size [Fig.~\ref{fig:acoustic_vs_r}(b)]. Although we considered only spherical particles here, it is expected that these features will be present in other geometries as well. 

The acoustic resonances are particularly interesting since they are completely absent in the single-fluid HDM, and experimental observation of the these peaks could serve as verification of the two-fluid model. To this end we analyzed different materials and different kinds of charge carriers. Here we saw that for the considered $p$-doped semiconductors with light and heavy holes, the damping was too high to discern any of the features of the two-fluid model. On the other hand, the intrinsic semiconductor particles that we studied, with thermally excited or laser-excited charge carriers, both have acoustic peaks in their spectra. 

\begin{acknowledgments}
We thank C. Cirac{\`i} for directing us to Ref.~\onlinecite{deceglia17a}.
We acknowledge support from the Danish Council for Independent Research (DFF 1323-00087). 
N.~A.~M. is a VILLUM Investigator supported by VILLUM FONDEN (grant No. 16498). Center for Nano Optics is financially supported by the University of Southern Denmark (SDU 2020 funding).
  \end{acknowledgments}

\appendix

\section{The susceptibility}
\label{dielectric}

We will show that the susceptibility is given by Eq.~(\ref{eq:chi2}). Starting from Eq.~(\ref{eq:chi1}), this can be rewritten in the following way by using the temporary variable $\kk'=-\kk-\q$

\begin{align}
\label{eq:chi7}
&\chi_i(\q,\w)=\frac{2 e^2}{\ep_0 q^2}\frac{1}{V}
\left(  \sum_{\kk}\frac{   f_i(\kk)}{E_i(\kk+\q)-E_i(\kk) -\hbar\w-i\eta}\right. \nonumber \\
&\qquad\qquad\qquad\left. -\sum_{\kk'}\frac{f_i(\kk')}{E_i(\kk')-E_i(\kk'+\q) -\hbar\w-i\eta}\right) , \nonumber \\
             &=\frac{4 e^2}{\ep_0 q^2}\frac{1}{V}
\sum_{\kk}f_i(\kk)\frac{E_i(\kk+\q)-E_i(\kk)}{\left(E_i(\kk+\q)-E_i(\kk)\right)^2 -\left(\hbar\w+i\eta\right)^2} . 
\end{align}
For holes the substitutions $f_i\to 1-f_i$ and $E_i\to E_v-E_i$ can be made in order to treat the electrons and holes on equal footing ($E_v$ is the valence band edge). However, this will leave Eq.~(\ref{eq:chi7}) unchanged. Next step is to take the limit $\q\to \mathbf{0}$ which allows for the series expansion
\begin{align}
\label{eq:chi8}
&\chi_i(\q,\w)=-\frac{4 e^2}{\ep_0 q^2}\frac{1}{V}
\sum_{\kk}f_i(\kk)\frac{E_i(\kk+\q)-E_i(\kk)}{\left(\hbar\w+i\eta\right)^2} \nonumber \\
&\qquad\times\left(1+\frac{\left(E_i(\kk+\q)-E_i(\kk)\right)^2}{\left(\hbar\w+i\eta\right)^2}+\ldots\right) .
\end{align}
Without loss of generality it is assumed that $\q=q\hat{\mathbf{z}}$ which means that
\begin{equation}
\label{eq:E_kq_Ek}
E_i(\kk+\q)-E_i(\kk)=\frac{\hbar^2}{2 m_i^*}\left(2 k_z q+q^2\right) ,
\end{equation}
and inserting this into $\chi_i(\q,\w)$ gives us
\begin{align}
\label{eq:chi9}
&\chi_i(q,\w)=-\frac{4 e^2}{\ep_0 q^2} \left(\frac{1}{(\hbar\w+i\eta)^2}\frac{\hbar^2q^2}{2 m_i^*} \frac{1}{V}\sum_{\kk}f_i(\kk)\right. \nonumber \\
&\quad\left.+\frac{1}{(\hbar\w+i\eta)^4}\frac{3\hbar^6q^4}{2 {m_i^*}^3} \frac{1}{V}\sum_{\kk}f_i(\kk)k_z^2+\cdots\right) .
\end{align}
where it has been taken into account that odd powers of $k_z$ cancel out.

To evaluate the $\kk$-sums in Eq.~(\ref{eq:chi9}) for light and heavy holes, we assume that $T=0\un{K}$ whereby the distribution becomes a step function. By using that the volume of a single state in $k$-space is $V_k=(2\pi)^3/V$, we find the first sum to be
\begin{align}
\label{eq:sum4}
&\frac{1}{V}\sum_{\kk}f_i(\kk) =\frac{1}{V V_k}\int \dif\kk f_i(\kk)\nonumber \\
&=\frac{4\pi}{V V_k}\int_0^{k_{Fi}} \dif k k^2=\frac{4\pi}{V V_k} \frac{k_{Fi}^3}{3} ,
\end{align}
which by definition also is equal to $n_i/2$. From this we also find the following simple relation
\begin{equation}
\label{eq:k_n}
k_{Fi}^3=3\pi^2 n_i .
\end{equation}
The second sum is given by
\begin{align}
\label{eq:sum5}
&\frac{1}{V}\sum_{\kk}f_i(\kk)k_z^2=\frac{1}{V V_k}\int_0^{k_{Fi}} \dif k k^4 \int_0^{\pi}\dif\theta\sin\theta\cos^2\theta\nonumber \\
& \times\int_0^{2\pi}\dif\phi=\frac{4\pi}{3 V V_k} \frac{k_{Fi}^5}{5}=\frac{k_{Fi}^2}{5}\frac{n_i}{2} .
\end{align}
The same results are obtained for laser-excited charge carriers, except that the Fermi levels $k_{F i}$ are for the quasi-equilibria that are assumed to be formed.

For the thermally excited intrinsic semiconductor, we will assume that the Fermi--Dirac distribution can be approximated by the Boltzmann distribution 
\begin{equation}
\label{eq:boltzmann}
f_e(E)=\frac{1}{\exp\left(\frac{E-E_F}{k_B T}\right)+1}\approx\exp\left(-\frac{E-E_F}{k_B T}\right) ,\nonumber
\end{equation}
which is reasonable for electrons whenever $E_c-E_F\gg k_B T$ where $E_c$ is the conduction band edge (a similar expression exist for holes). For electrons, the first sum becomes
\begin{align}
\label{eq:sum6}
&\frac{1}{V}\sum_{\kk}f_e(\kk) =\frac{4\pi}{V V_k}\int_0^{\infty} \dif k k^2f_e(k)\approx  \\
&\frac{2\pi}{V V_k}\left(\frac{2 m_e^*}{\hbar^2}\right)^{\frac{3}{2}}\int_{E_c}^{\infty} \dif E \sqrt{E-E_c}\exp\left(-\frac{E-E_F}{k_B T}\right)\nonumber ,
\end{align}
where it has been used that
\begin{equation}
\label{eq:E_k}
E=E_c+\frac{\hbar^2k^2}{2 m_e^*} .\nonumber
\end{equation}
We now introduce the variable $\rho=(E-E_F)/k_B T$ whereby the integral can be identified as a gamma function. With this, the sum is found to be
\begin{align}
\label{eq:sum8}
\frac{1}{V}\sum_{\kk}f_e(\kk) &\approx 
\frac{2\pi}{V V_k}\left(\frac{2 m_e^* k_B T}{\hbar^2}\right)^{\frac{3}{2}}\exp\left(\frac{E_F-E_c}{k_B T}\right)\frac{\sqrt{\pi}}{2} ,
\end{align}
which by definition also is equal to $n_e/2$. Using a similar method, the second sum is found to be
\begin{align}
\label{eq:sum9}
&\frac{1}{V}\sum_{\kk}f_e(\kk)k_z^2 \approx \frac{2\pi}{V V_k}\left(\frac{2 m_e^* k_B T}{\hbar^2}\right)^{\frac{5}{2}} \nonumber \\
&\times\exp\left(\frac{E_F-E_c}{k_B T}\right)\frac{1}{3}\Gamma\left(\frac{5}{2}\right)=\frac{m_e^* k_B T}{ \hbar^2}\frac{n_e}{2} .
\end{align}
The  sums for the holes can be found in the same way. 

Inserting the expressions of the sums into Eq.~(\ref{eq:chi9}), using $\eta^2\approx 0$ and defining $\gamma=2\eta/\hbar$ gives us Eq.~(\ref{eq:chi2}) where $\beta_i$ is given by either Eq.~(\ref{eq:beta1}) or (\ref{eq:beta2}) depending on the nature of the charge carriers.

\section{Charge carrier densities and Fermi wavenumbers}
\label{variables}

To find expressions for $k_{F lh}$ and $k_{F hh}$ for light and heavy holes, we will use the fact that the Fermi energy is the same for both kinds of holes
\begin{equation}
\label{eq:light_heavy1}
\frac{\hbar^2 k_{F lh}^2}{2 m_{lh}^*}=\frac{\hbar^2 k_{F hh}^2}{2 m_{hh}^*} .
\end{equation}
If we then use the relation between $n_i$ and $k_{Fi}$ from Eq.~(\ref{eq:k_n}) and assume complete ionization, $N_a=n_{lh}+n_{hh}$, we straight away get

\begin{equation}
\label{eq:k_light}
k_{Flh}=k_{Fhh}\sqrt{\frac{m_{lh}^*}{m_{hh}^*}}=\left[\frac{N_a 3\pi^2}{1+\left(\frac{m_{hh}^*}{m_{lh}^*}\right)^{\frac{3}{2}}}\right]^{\frac{1}{3}} .
\end{equation}

For laser-excited charge carriers, the energy density of a laser pulse that excites electrons from the valence band to the conduction band is given by
\begin{equation}
\label{eq:laser1}
u_{\rm pulse}=u_e+u_{lh}+u_{hh}+E_g n_e ,
\end{equation}
where $u_i$ is the energy density of the charge carrier type $i$ with respect to the band edge, and $e$, $lh$ and $hh$ are electrons, light holes and heavy holes, respectively. From Eq.~(\ref{eq:k_n}) and $E_i=\hbar^2 k^2/2 m_i^*$ we have $n_i=E_{Fi}^{3/2} (2 m_i^*)^{3/2}/ 3\pi^2\hbar^3$ and the energy densities are given by
\begin{equation}
\label{eq:laser2}
u_i=\frac{(2 m_i^*)^{\frac{3}{2}}}{2\pi^2\hbar^3}\int_0^{E_{Fi}}\dif E_i E_i^{\frac{3}{2}} =\frac{3\hbar^2}{10 m_i^*}\left( 3\pi^2\right)^{\frac{2}{3}} n_i^{\frac{5}{3}} .
\end{equation}
Inserting this into Eq.~(\ref{eq:laser1}) and using the following definition of the density-of-states hole mass\cite{sze} 
\begin{equation}
\label{eq:m_effective1}
m_h^*=\left({m^*_{lh}}^{\frac{3}{2}}+{m^*_{hh}}^{\frac{3}{2}}\right)^{\frac{2}{3}} ,
\end{equation}
together with charge conservation $n_e=n_h=n_{lh}+n_{hh}$ and Eq.~(\ref{eq:light_heavy1}) we obtain
\begin{equation}
\label{eq:n_laser}
u_{\rm pulse}=\frac{3 \hbar^2}{10}\left(3 \pi^2\right)^{\frac{2}{3}}\left(\frac{1}{m_e^*}+\frac{1}{m_h^*}\right)n_e^{\frac{5}{3}}+E_g n_e .     
\end{equation}
From this expression, the density of electrons can be found using numerical tools, and $k_{Fi}$ is found using Eq.~(\ref{eq:k_n}). 

For thermally excited charge carriers in an intrinsic semiconductor, the charge carrier densities are given by\cite{sze}
\begin{equation}
\label{eq:n_therm}
n_e\!=\!n_h\!=2\left(\frac{2\pi k_B T}{h^2}\right)^{\frac{3}{2}}\! {m^*_e}^{\frac{3}{4}} {m^*_h}^{\frac{3}{4}}\exp\left(\frac{-E_g}{2 k_B T}\right) ,
\end{equation}
where $E_g$ is the band gap. Here it is assumed that the Boltzmann distribution can be used for the electrons.

For both laser-excited and thermally excited charge carriers the density-of-states effective hole mass is given by Eq.~(\ref{eq:m_effective1}), while the conductivity effective hole mass is given by\cite{spitzer57}
\begin{equation}
\label{eq:m_effective2}
m_{h,{\rm cond}}^*=\frac{{m^*_{lh}}^{\frac{3}{2}}+{m^*_{hh}}^{\frac{3}{2}}}{{m^*_{lh}}^{\frac{1}{2}}+{m^*_{hh}}^{\frac{1}{2}}} .
\end{equation}

\section{Matrix notation}
\label{matrix}

Here, we rewrite the two-fluid equations~(\ref{eq:hydro2}) in a matrix notation

\begin{subequations}
\begin{equation}
\hat{\cal L}\E=-i\w\mu_0 \begin{pmatrix}
1\\1
\end{pmatrix} \begin{pmatrix}
\J_a\\\J_b
\end{pmatrix} ,
\end{equation}
\begin{align}
\label{eq:constitutive-eq-matrix}
&\overbrace{\begin{pmatrix}
\frac{\beta_a^2}{\w_a^2} & 0\\
0& \frac{\beta_b^2}{\w_b^2}
\end{pmatrix}}^{\equiv \bm{\xi}^2 }\nabla (\Div) 
\begin{pmatrix}
\J_a\\ \J_b
\end{pmatrix}  \\
&+k^2\underbrace{\begin{pmatrix}
\frac{c^2}{\w_a^2}& 0\\
0& \frac{c^2}{\w_b^2}
\end{pmatrix}\left[
{\mathbf I}+\begin{pmatrix}
\frac{i\gamma_a}{\w} & 0\\
0 & \frac{i\gamma_b}{\w}
\end{pmatrix}\right]}_{\equiv {\mathbf \Lambda}^2 } 
\begin{pmatrix}
\J_a\\ \J_b
\end{pmatrix}
=i\w\ep_0 \begin{pmatrix}
1\\1
\end{pmatrix}\E ,\nonumber
\end{align}
\end{subequations}
where $\hat{\cal L}=-\Curl\Curl+ \ep_{\infty}k^2$ with $k=\omega/c$, while ${\mathbf I}$ is a $2\times 2$ identity matrix. Next, we follow a trick developed in Ref.~\onlinecite{toscano13}, where one acts with $\hat{\cal L}$ on the constitutive equation~(\ref{eq:constitutive-eq-matrix}). At first sight, this generates less appealing 4th-order derivatives, but the curl of any gradient field vanishes, and we are eventually left with only 2nd-order derivatives, i.e.

\begin{equation}
\label{eq:matrix1}
\Big[\ep_{\infty}\bm{\xi}^2\nabla (\Div) -{\mathbf \Lambda}^2\Curl\Curl - {\mathbf M}\Big]
\begin{pmatrix}
\J_a\\ \J_b
\end{pmatrix}
=0 ,
\end{equation}
where ${\mathbf M}\equiv\tilde{\mathbf 1}   -\ep_{\infty}k^2{\mathbf \Lambda}^2$ and $\tilde{\mathbf 1}$ is a $2\times 2$ all-ones matrix. While $\bm{\xi}$ and ${\mathbf \Lambda}$ are diagonal matrices, ${\mathbf M}$ has non-zero off-diagonal elements and the two currents are consequently coupled. The coupling originates from a mutual interaction through common electromagnetic fields (which we have integrated out). 

To find the uncoupled, homogeneous equations for the normal modes, we take either the curl or the divergence of Eq.~(\ref{eq:matrix1}) and obtain the following equations
\begin{subequations}
\begin{align}
\label{eq:matrix2t}
\Big[{\mathbf \Lambda}^2\nabla^2 - {\mathbf M}\Big]
\Curl\begin{pmatrix}
\J_a\\ \J_b
\end{pmatrix}
=0 ,\\
\label{eq:matrix2l}
\Big[\ep_{\infty}\bm{\xi}^2\nabla^2 - {\mathbf M}\Big]
\Div\begin{pmatrix}
\J_a\\ \J_b
\end{pmatrix}
=0 ,
\end{align}
\end{subequations}
where it is used that $\nabla^2=\nabla(\Div)-\Curl\Curl$. Next, the linear relations between $\{J_a,J_b\}$ and $\{J_1,J_2\}$ given in Eqs.~(\ref{eq:j1j2}) are introduced for both the transversal fields (the curl-equation) and the longitudinal fields (the divergence-equation) which gives us
\begin{subequations}
\begin{align}
\label{eq:matrix3t}
\Big[{\mathbf \Lambda}^2\mathbf{K}_T\nabla^2 - {\mathbf M}\mathbf{K}_T\Big]
\Curl\begin{pmatrix}
\J_1\\ \J_2
\end{pmatrix}
=0 ,\\
\label{eq:matrix3l}
\Big[\ep_{\infty}\bm{\xi}^2\mathbf{K}_L\nabla^2 - {\mathbf M}\mathbf{K}_L\Big]
\Div\begin{pmatrix}
\J_1\\ \J_2
\end{pmatrix}
=0 ,
\end{align}
\end{subequations}
where 
\begin{equation}
\mathbf{K}_z=
\begin{pmatrix}
a_1^z & a_2^z \\ b_1^z & b_2^z
\end{pmatrix} ,\nonumber
\end{equation}
with $z=T,L$. If we then \emph{require} that $\J_1$ and $\J_2$ are uncoupled for both the transversal and the longitudinal fields, the $2\times 2$ non-diagonal matrices can be treated as $4\times 4$ diagonal matrices. In other words, we obtain 8 homogeneous equations in total: for both curl and divergence we get two for both $\J_1$ and $\J_2$. These are the Boardman equations written explicitly in section~\ref{bulk}. The fact that there are two equations for every $\Curl\J_j$ and $\Div\J_j$ can be used to find the coefficients $a_j^z$ and $b_j^z$ which so far have been undetermined.

\section{Linear equations}
\label{linear}

When applying the boundary conditions $\Delta\E_{\parallel}=0$, $\Delta\B_{\parallel}=0$, $\J_{a,\perp}=0$ and $\J_{b,\perp}=0$ to the electrical fields in Eqs.~(\ref{eq:Ei}), (\ref{eq:Er}) and (\ref{eq:Et}), the following system of linear equations is obtained

\begin{subequations}
\label{eq:linear_equations1}
\begin{align}
\label{eq:linear1}
&-a_l^r h_l^{(1)}(x_D)+a_l^t j_l(x_T)=j_l(x_D) ,\\
\label{eq:linear2}
&-a_l^r\lbrack x_D h^{(1)}_l(x_D)\rbrack^\prime+a_l^t\lbrack x_T j_l(x_T)\rbrack^\prime = \lbrack x_D j_l(x_D)\rbrack^\prime , \\
\label{eq:linear3}
&-b_l^r\frac{\lbrack x_D h^{(1)}_l(x_D)\rbrack^\prime}{k_D}+b_l^t\frac{\lbrack x_T j_l(x_T)\rbrack^\prime}{k_T} \nonumber \\
&\qquad+i c_{1l}^t j_l(x_1)+i c_{2l}^t j_l(x_2)= \frac{\lbrack x_D j_l(x_D)\rbrack^\prime}{k_D} , \\
\label{eq:linear4}
&-b_l^r x_D h_l^{(1)}(x_D)+b_l^t x_T j_l(x_T)=x_D j_l(x_D) , \\
\label{eq:linear5}
&-i b_l^t \frac{l (l+1)}{x_T} j_l(x_T)
+c_{1l}^t j_l'(x_1) k_{L,1}\left(1+\frac{\beta_a^2\ep_{\infty} k_{L,1}^2}{\w_a^2(1+\alpha_1)}\right) \nonumber \\
&\qquad+c_{2l}^t j_l'(x_2) k_{L,2}\left(1+\frac{\beta_a^2\ep_{\infty} k_{L,2}^2}{\w_a^2(1+\alpha_2)}\right)=0  , \\
\label{eq:linear6}
&-i b_l^t \frac{l (l+1)}{x_T} j_l(x_T)
+c_{1l}^t j_l'(x_1) k_{L,1}\left(1+\frac{\beta_b^2\ep_{\infty} k_{L,1}^2}{\w_b^2(1+\alpha_1^{-1})}\right) \nonumber \\
&\qquad+c_{2l}^t j_l'(x_2) k_{L,2}\left(1+\frac{\beta_b^2\ep_{\infty} k_{L,2}^2}{\w_b^2(1+\alpha_2^{-1})}\right)=0 ,
\end{align}
\end{subequations}
which directly allows us to find $a_l^r$ and $b_l^r$.

\end{document}